\documentclass[english,aps,superscriptaddress,preprintnumbers,reprint,footinbib,amsmath,amssymb,prb]{revtex4-2}
\usepackage[latin9]{inputenc}
\usepackage{color}
\usepackage{amsmath}
\usepackage{amssymb}
\usepackage{graphicx}
\usepackage{esint}

\makeatletter

\@ifundefined{textcolor}{}{%
 \definecolor{BLACK}{gray}{0}
 \definecolor{WHITE}{gray}{1}
 \definecolor{RED}{rgb}{1,0,0}
 \definecolor{GREEN}{rgb}{0,1,0}
 \definecolor{BLUE}{rgb}{0,0,1}
 \definecolor{CYAN}{cmyk}{1,0,0,0}
 \definecolor{MAGENTA}{cmyk}{0,1,0,0}
 \definecolor{YELLOW}{cmyk}{0,0,1,0}
}

\usepackage{aecompl}

\usepackage{epsfig}\usepackage{dcolumn}\usepackage{bm}
\usepackage{babel}

\usepackage{hyperref}
\hypersetup{
    colorlinks=true,
    linkcolor=blue,
    urlcolor=blue,
    citecolor=blue
        }

\makeatother

\usepackage{babel}
\begin{document}
\title{Fractionalization of {Majorana-Ising-type quasiparticle
excitations}}
\author{J.E. Sanches}
\email[corresponding author: ]{jose.sanches@unesp.br}

\affiliation{São Paulo State University (Unesp), School of Engineering, Department
of Physics and Chemistry, 15385-000, Ilha Solteira-SP, Brazil}
\author{L.T. Lustosa}
\affiliation{São Paulo State University (Unesp), School of Engineering, Department
of Physics and Chemistry, 15385-000, Ilha Solteira-SP, Brazil}
\author{L.S. Ricco}
\affiliation{Science Institute, University of Iceland, Dunhagi-3, IS-107, Reykjavik,
Iceland}
\author{I.A. Shelykh}
\affiliation{Science Institute, University of Iceland, Dunhagi-3, IS-107, Reykjavik,
Iceland}
\affiliation{Abrikosov Center for Theoretical Physics, MIPT, Dolgoprudnyi, Moscow
Region 141701, Russia}
\author{M. de Souza}
\affiliation{São Paulo State University (Unesp), IGCE, Department of Physics, 13506-970,
Rio Claro-SP, Brazil}
\author{M.S. Figueira}
\affiliation{Instituto de Física, Universidade Federal Fluminense, 24210-340, Niterói,
Rio de Janeiro, Brazil}
\author{A.C. Seridonio}
\email[corresponding author: ]{antonio.seridonio@unesp.br}

\affiliation{São Paulo State University (Unesp), School of Engineering, Department
of Physics and Chemistry, 15385-000, Ilha Solteira-SP, Brazil}
\begin{abstract}
We theoretically investigate the spectral properties of a quantum
impurity (QI) hosting the here proposed {Majorana-Ising-type
quasiparticle (MIQ) excitation}. It arises from the coupling between
a finite topological {superconductor} (TSC) based
on a chain of magnetic adatoms-superconducting hybrid system and an
integer large spin $S$ flanking the QI. Noteworthy, the spin $S$
couples to the QI via the Ising-type exchange interaction.{{}
As the Majorana zero-modes (MZMs) at the edges of the TSC chain are
overlapped, we counterintuitively find a regime wherein the Ising
term modulates the localization of a fractionalized and resonant MZM
at the QI site. Interestingly enough, the fermionic nature of this
state is revealed as purely of electron tunneling-type and most astonishingly,
it has the Andreev conductance completely null in its birth. Therefore,
we find that a resonant edge state appears as a zero-mode and discuss
it in terms of a }\textit{{poor man's Majorana}}{{[}Nature
614, 445 (2023){]}.}
\end{abstract}
\maketitle

\section{Introduction}

Majorana fermions are peculiar particles equal to their own antiparticles
described by real solutions of the Dirac equation{\citep{Majorana-1937}}.
In {condensed matter Physics}, such fermions rise
as quasiparticle excitations usually quoted as Majorana zero-modes
(MZMs), which are found attached to the boundaries of topological
superconductors (TSCs){\citep{Marra-2022,Baranger-2011,Leakage-2014,Brouwer-2011,DasSarma-2009,Yuval-Oreg-2022,Oppen-2014,Oppen-2010,Oppen-2013,MacDonald-2013,Klinovaja-2013,Yazdani-2013,Fu-Kane-2009,Fu-Kane-2008,M.-Franz-2013,Nagaosa-2013,Pascal-Simon-2013,Potter-Lee-2012,S-CZhang-2011,Beenakker-2015,Fujimoto-2016,YoichiAndo-2017,Yuval-Oreg-2019,Alicea-2012,Beenakker-2013,Flensberg-2021,Klinovaja-2021,C.Marcus-2016,Mourik-2012,C.Marcus-2017}}.
Astonishingly, since the theoretical Kitaev seminal proposal of \textit{p-wave}
superconductivity{\citep{Kitaev-2001}}, MZMs are notably coveted
due to their attribution as building-blocks for the highly pursued
fault-tolerant topological quantum computing. Thereafter, in the last
years, such excitations have received astounding focus from both communities
working with quantum science and technology.

{Interestingly enough, theoretical predictions point
out that the fractional zero-bias peak (ZBP) in transport evaluations
through quantum dots, which is given by the conductance $\mathcal{G}_{\text{{Total}}}(0)=\frac{e^{2}}{2h}$\citep{Marra-2022,Baranger-2011,Leakage-2014},
could have its origin from both the system topological nontrivial
regime, where two MZMs emerge spatially far apart at the edges of
a TSC{\citep{Baranger-2011,Leakage-2014}}, as well as in the corresponding
trivial, which could exhibit, for instance, Andreev bound states (ABSs)\cite{Nonlocality_Majorana,Ricco_2019}.
In this regard, we highlight Ref.\citep{AndreevVersusMajorana}, which
by treating the TSC within the theoretical framework of the Oreg-Lutchyn
Hamiltonian\cite{Oppen-2010}, allows a detailed and systematic analysis
on the formation of MZMs versus ABSs issue. In parallel, effective
models\cite{Baranger-2011,Nonlocality_Majorana,IsoG}, despite their
simplifications, are used to capture, with a quite good accuracy,
the corresponding low-energy Physics encoded by models such as that
in Ref.\cite{Oppen-2010} and the Kitaev Hamiltonian\citep{Kitaev-2001}. }

\begin{figure}[t]
\centering\includegraphics[width=1\columnwidth]{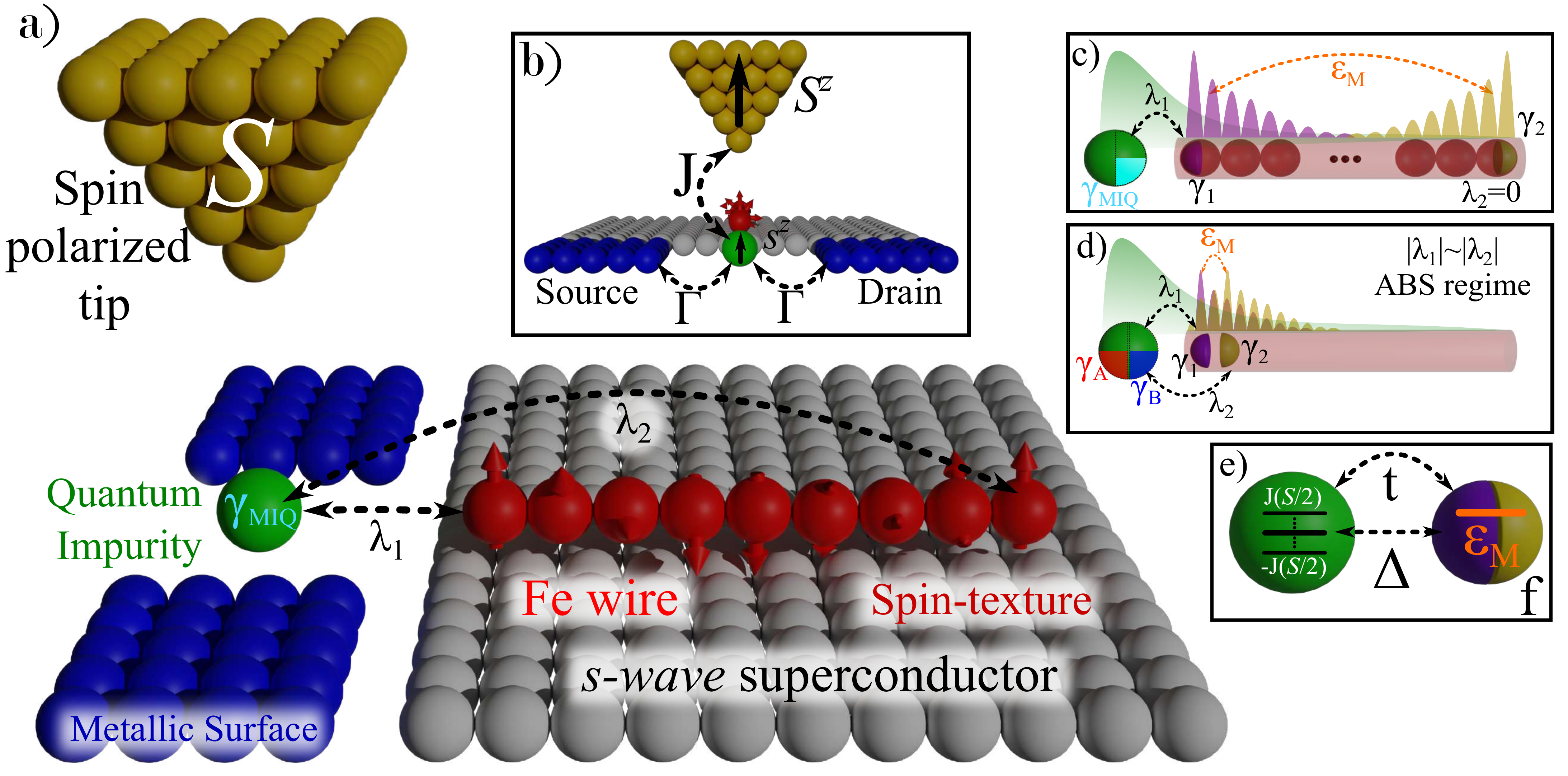} \caption{\label{fig:Fig1} (a) Proposed device expected to show a {Majorana-Ising-type
quasiparticle (MIQ) excitation} $\gamma_{\text{{MIQ}}}$ localized
at the quantum impurity (QI) site. {The $\gamma_{\text{{MIQ}}}$
} leads to a zero-bias conductance peak $\mathcal{G}_{\text{{MIQ}}}(0)=\frac{e^{2}}{4h}$
due to the QI placed between source and drain leads, but the total
conductance is still $\mathcal{G}_{\text{{Total}}}(0)=\frac{e^{2}}{2h}.$
This occurs once a genuine electron tunneling process is present and
there is a complete lack of local Andreev reflection. It can be performed
by considering the QI simultaneously coupled to a spin-polarized tip
with an integer large spin $S$ via an Ising-type exchange interaction
$J$ and with {the hopping terms $\lambda_{1(2)}$
(being $\lambda_{2}=0$ for this case)} to a helical spin-texture
chain hosted by an \textit{s}-wave superconductor. (b) Side view of
panel (a) wherein the QI-lead coupling $\Gamma$ and Ising-type exchange
interaction $J$ for the spin-polarized tip and QI appear highlighted.
(c) Pictorial scheme of panel (b), which effectively consists of a
topological superconductor (TSC), where $\epsilon_{M}$ represents
the overlap between the Majorana zero-modes (MZMs) $\gamma_{1}$ and
$\gamma_{2}$ at edges, while the QI shows the MIQ $\gamma_{\text{{MIQ}}}.$
The spatial distributions of the wave functions for the MZMs and QI
are also illustrated. The factor $\frac{1}{4}$ in $\mathcal{G}_{\text{{MIQ}}}$
has correspondence to the $\frac{1}{4}$ from the volume for the sphere
depicted to represent the quasiparticle $\gamma_{\text{{MIQ}}},$
in contrast to the ratio $\frac{1}{2}$ for the typical volume of
the MZMs $\gamma_{1}$ and $\gamma_{2}.${{} (d) Andreev
bound state (ABS) regime, obtained from the effective model with $|\lambda_{1}|\sim|\lambda_{2}|$\cite{Nonlocality_Majorana}.}
(e) These MZMs mimic a delocalized fermionic site $f,$ wherein $\epsilon_{M}$
plays the role of its energy level, while the QI has $2S+1$ levels
ranging from $-J(S/2)$ to $+J(S/2).$ In this scenario, such quantum
dots constitute a Kitaev dimer, i.e., with hopping $t$ and pairing
$\triangle$ of the \textit{p-wave} Cooper pair split into the QI
and $f$ orbitals.}
\end{figure}

{Back to the issue of the topological nontrivial regime,
not less important, once an ordinary fermion can be decomposed into
two MZMs, is the amplitude $1/2$ in $\mathcal{G}_{\text{{Total}}}(0),$
the hallmark of the fractionalization of the quantum of conductance
$\frac{e^{2}}{h},$ thus giving rise to the concept of fractionalized
electronic zero-frequency spectral weight, which indeed reveals, the
MZM ``half-fermionic'' nature\cite{Leakage-2014}.} This aforementioned
fingerprint is expected to show up in engineered platforms that combine
conventional \textit{s-wave} superconductivity and spin-texture {[}see
Refs.{\citep{Yazdani-2014,J.Franke-2015,Pawlak_2016}} and Fig.\ref{fig:Fig1}(a){]}.
As aftermath, the \textit{p-wave} superconductivity becomes feasible,
thus allowing the experimental realization of the spinless Kitaev
wire, which is indeed, a TSC in 1D{\citep{Kitaev-2001,Alicea-2012,Colloquium-Franz-2015,Aguado-2017,Yuval-Oreg-2022,Flensberg-2012,Tewari-2013,Flensberg-2021}}.
{For such an accomplishment, we highlight two practical
recipes based on the Oreg-Lutchyn proposal\cite{Oppen-2010}, which
have the following ingredients: }(i) a semiconducting nanowire, with
strong spin-orbit coupling (SOC) and under a magnetic field, should
be deposited on top of an \textit{s-wave} superconductor{\citep{C.Marcus-2016,Mourik-2012,C.Marcus-2017,Alicea-2012,Colloquium-Franz-2015,Aguado-2017,DasSarma-2010}}
or (ii) a linear chain of magnetic adatoms with exchange interactions
should be hosted by an \textit{s-wave} superconductor with strong
SOC{\citep{Pawlak_2016,J.Franke-2015,Beenakker-2011,Yazdani-2013,Loss-2013,Pascal-Simon-2013,Yazdani-2014,Yazdani-2017,Nitta-2019,Ernst-Meyer-2019,Yazdani-2021,swaveCoupling}}.
In both the situations, the \textit{s-wave} superconductors with singlet
Cooper pairing lead to the so-called superconducting proximity effect,
which is pivotal to carry forward the superconducting (SC) character
into such manufactured Kitaev wires. Thus, the Zeeman field from the
previous recipes (i) and (ii), together with the magnified SOC from
such quantum materials, establish a synergy that stabilizes the system
spin-texture. Consequently, the triplet Cooper pairing for the \textit{p-wave}
superconductivity, as well.

Particularly in the topological nontrivial regime of such setups,
these Majorana quasiparticle excitations emerge ideally, i.e., as
MZMs decoupled from each other and localized on the boundaries of
the TSC. Due to this decoupling from their environment, MZMs are regarded
robust against perturbations, once they are topologically protected
by the SC gap. Thus, MZMs become promising candidates for a quantum
computing free of the decoherence phenomenon\citep{DasSarma-2008,Nayak-2015}.
However, perfectly far apart MZMs are hypothetical objects, since
they are reliable solely in infinite-size systems and in real experiments,
the quantum wires are finite. As a result, these end MZMs within a
finite length overlap with each other and inevitably, a fermionic
mode with a finite energy emerges instead.

To overcome the aforementioned challenge, in this work, we find as
route the fractionalization of ordinary MZMs, in particular, those
found at a quantum impurity (QI) site coupled to one edge of a finite
TSC in 1D. To this end, we should take into account the Ising exchange
interaction between an integer large spin and such a QI. This setup
corresponds to Figs.\ref{fig:Fig1}(a) and (b), and it contains the
ordinary MZMs $\gamma_{1}$ and $\gamma_{2}$ placed at the edges
of a short TSC wire. To better understand our findings, we propose
to view the MZMs as sketched in Fig.\ref{fig:Fig1}(c), where these
objects appear symbolized by calottes (half-spheres). We clarify that
the employment of such a pictorial representation for the MZMs aims
to explain diagrammatically the electron fractionalization into them,
as well as the MZM fractionalization itself here observed. These calottes
belong to a delocalized sphere cut in half, with each part placed
at the TSC edges. This cartoon is very useful and it has the purpose
of emulating the nonlocal nature of the fermionic state composed by
these MZMs, which are found spatially far apart. To best of our knowledge,
the Majorana zero-frequency spectral weight, in particular for a QI
coupled to an infinite TSC, is given by the unity when the leaking
of the MZM $\gamma_{1}$ into a quantum dot occurs{\citep{Leakage-2014}}.
This unity corresponds to a calotte, which is a half-electron state
that contributes to the conductance $\mathcal{G}_{\text{{MZM}}}(0)=\frac{e^{2}}{2h},$
as expected{\citep{Baranger-2011,Leakage-2014}}. Particularly for
a finite TSC, we define as the system sweet spot {[}see Fig.\ref{fig:Fig3}b)-I{]}
a special configuration, in which a peculiar Ising exchange interaction
allows us to observe{{} a fractionalized MZM quasiparticle
excitation $\gamma_{\text{{MIQ}}}$} at the QI, namely, {the
called by us as Majorana-Ising-type quasiparticle (MIQ) excitation
} {[}Figs.\ref{fig:Fig1}(a) and (c){]}. {This quasiparticle
excitation} can be viewed as the half-calotte within the cartoon representation
of the QI state {[}see also Fig.\ref{fig:Fig1}(c){]}. Such a ratio
symbolizes a novel MZM-type excitation in the presence of finite TSCs,
in which technically speaking, it is identified by a Majorana zero-frequency
spectral weight equals to half. Equivalently, the same amount corresponds
to $\frac{1}{4}$ of the entire QI electronic state. In contrast,
it leads to $\mathcal{G}_{\text{{MIQ}}}(0)=\frac{e^{2}}{4h}$ as it
should be. Interestingly enough, solely in one of the Majorana densities
of states (DOSs) of the QI, the MIQ becomes evident as a resonant
mode localized at $\omega=0.$ Counterintuitively, the other Majorana
DOS of the QI instead of exhibiting a resonant state, reveals a Majorana
zero-frequency spectral weight with a valley, but presenting the same
magnitude of the resonant Majorana fermion. In this manner, we can
safely state that this {novel MZM-type quasiparticle
excitation, which for simplicity, we just call by MIQ from now on,}
is then found at the QI. In this situation, we demonstrate that the
emergence of such {a quasiparticle excitation yields
a zero-bias local Andreev conductance entirely null}, with only normal
electronic contribution to the total conductance.

\section{The Model}

The effective system Hamiltonian that corresponds to the proposed
setup presented in Fig.\ref{fig:Fig1}(a) can be expressed as:
\begin{eqnarray}
\mathcal{H} & = & \sum_{\alpha k\sigma}\varepsilon_{\alpha k}c_{\alpha k\sigma}^{\dagger}c_{\alpha k\sigma}+\sum_{\sigma}\varepsilon_{\sigma}d_{\sigma}^{\dagger}d_{\sigma}+i\varepsilon_{M}\gamma_{1}\gamma_{2}\nonumber \\
 & + & \left(\frac{\Gamma}{2\pi\rho_{0}}\right)^{1/2}\sum_{\alpha k\sigma}(c_{\alpha k\sigma}^{\dagger}d_{\sigma}+\text{{H.c.}})+Js^{z}S^{z}\nonumber \\
 & + & \lambda_{1}(d_{\uparrow}-d_{\uparrow}^{\dagger})\gamma_{1}{{+\lambda_{2}(d_{\uparrow}+d_{\uparrow}^{\dagger})\gamma_{2}},}\label{eq:Htotal}
\end{eqnarray}
where the operator $c_{\alpha k\sigma}^{\dagger}(c_{\alpha k\sigma})$
describes the creation (annihilation) of an electron with momentum
$k$, spin-$z$ $\sigma=\pm1,$ energy $\varepsilon_{\alpha k}=\varepsilon_{k}-\mu_{\alpha}$
for the metallic lead $\alpha=[\text{{Source}},\text{{Drain}}]$ in
terms of the single-particle energy $\varepsilon_{k}$ and chemical
potential $\mu_{\alpha},$ while $d_{\sigma}^{\dagger}(d_{\sigma})$
stands for the electrons at the QI site in which $\varepsilon_{\sigma}$
represents their energy levels per spin. To connect the QI to the
metallic leads and a large spin $S$ as well, we should consider the
QI-lead coupling $\Gamma=2\pi v^{2}\rho_{0},$ which is determined
by the QI-lead hopping term $v$ and lead DOS $\rho_{0},$ in parallel
to the Ising-type exchange interaction $J$ {[}Fig.\ref{fig:Fig1}(b){]}.
This Ising Hamiltonian involves the components $s^{z}$ and $S^{z}$
of the QI $(s=1/2)$ and $S,$ respectively, wherein the latter could
be well-represented, within an experimental framework, by a spin-polarized
tip {[}Ref.\citep{Spintips} and Figs.\ref{fig:Fig1}(a)-(b){]}. The
emergence of MZMs at the TSC wire edges are accounted for $\gamma_{1}$
and $\gamma_{2},$ with $\varepsilon_{M}$ as the overlap parameter.
Finally, {$\lambda_{1}$ and $\lambda_{2}$ couple
the spin-up channel of the QI to $\gamma_{1}$ and $\gamma_{2}$,
respectively {[}Figs.\ref{fig:Fig1}(a) and (c){]}}. Additionally,
for the sake of simplicity, we consider that the spin-down degree
is decoupled from the TSC and obeys the single impurity Anderson model
(SIAM){\citep{Anderson-1961}}. Thus, we assume that the spin component
of the QI that couples to the Kitaev wire is $\sigma=+1,$which is,
as can be viewed in Fig.\ref{fig:Fig1}(a), the same spin direction
assumed for the edges of the magnetic chain of adatoms. This means
that the spin-flips of the QI electron injected into the TSC and vice-versa
are prevented. Therefore, the spin-up degree of the QI is the unique
to perceive the TSC. Different spin-textures on the TSC edge where
$\gamma_{1}$ is found\citep{Aguado-non}, which would allow the mixing
of the spin degrees of freedom, will be addressed elsewhere and do
not belong to the current analysis.{{} Additionally,
as we assume the intrinsic Zeeman splitting $\varepsilon_{\downarrow}-\varepsilon_{\uparrow}=V_{Z}$
of the magnetic chain as the largest energy scale, the spin-down has
no influence in the phenomenon here reported, once the corresponding
electronic occupation is empty.}{{} However, even
with the present assumption, we are free to demonstrate that both
these spin components become influenced by $S$ and the TSC that mimics
an effective quantum dot tunnel and Andreev-coupled to the QI {[}Fig.\ref{fig:Fig1}(e),\cite{Flensberg_2012(Poor),Kouwenhoven_2023}{]}.}

To this end, we call the attention that the Majorana and Ising terms
of Eq.(\ref{eq:Htotal}) should be conveniently rewritten to access
the system underlying Physics: (i) the Ising term turns straightforwardly
into
\begin{equation}
Js^{z}S^{z}=\frac{J}{2}\sum_{m\sigma}\sigma md_{\sigma}^{\dagger}d_{\sigma}|m\rangle\langle m|,\label{eq:HIsing}
\end{equation}
due to the standard expansions $s^{z}=\frac{1}{2}\sum_{\sigma}\sigma d_{\sigma}^{\dagger}d_{\sigma}$
and $S^{z}=\sum_{m}m|m\rangle\langle m|$ with $m=[-S,-S+1,...,S-1,S]$
for the QI and large spin, respectively. This means that each spin
channel in the QI acquires a multi-level structure split into $2S+1$
energies ranging from $-J(S/2)$ to $+J(S/2).$ {As
a matter of fact, the TSC alters this feature quite a bit for the
channel $\sigma=+1;$} and (ii) it is imperative to evoke that the
MZMs are made by the electron $(f^{\dagger})$ and hole $(f)$ of
a regular Dirac fermion delocalized over the TSC edges, which lead
to $\gamma_{1}=\frac{1}{\sqrt{2}}(f^{\dagger}+f)$ and $\gamma_{2}=\frac{i}{\sqrt{2}}(f^{\dagger}-f).$
In this picture, $\varepsilon_{M}$ plays the role of the energy level
related to the electronic occupation $f^{\dagger}f$ and the QI is
indeed hybridized with $f,$ as mentioned earlier, via the hopping
$t$ and the superconducting pairing $\Delta.$ {In
summary, by considering the parameterization $\lambda_{1}=(t+\Delta)/\sqrt{2}$
and $\lambda_{2}=i(\Delta-t)/\sqrt{2}$ {[}Fig.\ref{fig:Fig1}(a){]},
we simply find the Kitaev dimer composed by $d_{\uparrow}$ and $f$
orbital sites\cite{Flensberg_2012(Poor),Kouwenhoven_2023}:
\begin{align}
i\varepsilon_{M}\gamma_{1}\gamma_{2}+\varepsilon_{\uparrow}d_{\uparrow}^{\dagger}d_{\uparrow}+\lambda_{1}(d_{\uparrow}-d_{\uparrow}^{\dagger})\gamma_{1}+{{\lambda_{2}(d_{\uparrow}+d_{\uparrow}^{\dagger})\gamma_{2}}}\nonumber \\
=\varepsilon_{M}(f^{\dagger}f-\frac{1}{2})+\varepsilon_{\uparrow}d_{\uparrow}^{\dagger}d_{\uparrow}+(td_{\uparrow}f^{\dagger}+\Delta d_{\uparrow}f+\text{H.c.}).\label{eq:HMZMs}
\end{align}
}Similarly, the spin-up channel of the QI can be also decomposed in
MZMs{\citep{Leakage-2014}}, which we label by $\gamma_{A}$ and
$\gamma_{B},$ i.e.,
\begin{equation}
d_{\uparrow}=\frac{1}{\sqrt{2}}(\gamma_{A}+i\gamma_{B}).\label{eq:Dup}
\end{equation}
Based on Eq.(\ref{eq:Dup}), one can compute the Majorana zero-frequency
spectral weights for $\gamma_{A}$ and $\gamma_{B},$ respectively.
{These quantities reveal that in the }{poor
man's Majorana} regime ($J=0,$ $\Delta=t$ and
$\varepsilon_{M}=\varepsilon_{\uparrow}=0$)\cite{Flensberg_2012(Poor),Kouwenhoven_2023}
for the Kitaev dimer, that at the QI site, one spectral weight shows
a unitary amplitude for a resonant zero-mode, while the other is completely
null at zero-energy. It means that an isolated MZM is found at the
QI. Analogously, such a feature is also observed in $f$\citep{Kitaev-2001,Leakage-2014}.
This results, according to Eq.(\ref{eq:HMZMs}) given by $i\sqrt{2}\lambda_{1}\gamma_{B}\gamma_{1}$,
into{{} two isolated MZMs spatially placed at $d_{\uparrow}$
and $f$ orbital sites,} namely, $\gamma_{A}$ and $\gamma_{2},$
respectively.{{} Although $f$ is already nonlocal
and split over the TSC edges, it should be understood as an orbital
site in the framework of the Kitaev dimer, thus $\gamma_{2},$ within
such a context, is placed there.} For the trivial case $(\varepsilon_{M}\neq0)$,
the two spectral weights for $\gamma_{A}$ and $\gamma_{B}$ attain
to unity and then, two resonant zero-modes emerge at the QI site.
Thus, Eq.(\ref{eq:Dup}) is extremely clarifying, once it points out
the possibility of having within the QI, in the presence of $\varepsilon_{M}$
and $J,$ the isolation of the here proposed MIQ. Additionally, as
one can notice, the spin down channel always shows the trivial case,
due to its decoupling from the TSC. For completeness, the MZMs for
$d_{\downarrow}$ we label by $\gamma_{C}$ and $\gamma_{D}.$ {To
conclude, we are aware that the QI could also be coupled to the}\textit{{{}
s-wave}}{{} platform of Fig.\ref{fig:Fig1}(a) as
in Ref.\citep{swaveCoupling} by the term $\mathcal{H}_{s-QI}\sim-\Gamma_{s}(d_{\uparrow}^{\dagger}d_{\downarrow}^{\dagger}+\text{{H.c.}})$\citep{swaveCoupling,IsoG}
in the limit $\Delta_{\text{{SC}}}\rightarrow\infty$ (infinite superconducting
gap standard approximation), where $\Gamma_{s}$ is the s-wave-QI
coupling. Nevertheless, differently, we do not consider the on top
geometry of Ref.\citep{swaveCoupling} and assumes intrinsic Zeeman
splitting $\varepsilon_{\downarrow}-\varepsilon_{\uparrow}=V_{Z}$
of the magnetic chain extremely high, in order to rid off the spin-down
component. As a result, we expect that both features suppress $\mathcal{H}_{s-QI}$
and let the exploration to elsewhere when $V_{Z}$ is not found magnified.}

\section{quantum transport and Green's functions }

In this section, our goal is the analytical evaluation of the total
conductance through the QI device depicted in Figs.\ref{fig:Fig1}(a)-(b).
As a matter of fact, solely the spin-up channel contributes to the
conductance, once the spin-down is energetically inaccessible as previously
stated.{{} As our goal is the transport determination
around the energy of the MZM, only the bias-energy between source
and drain leads $\left|\text{{eV}}\right|\ll\Delta_{\text{{SC}}}\rightarrow\infty$
(infinite superconducting gap standard approximation) is accounted
for in the derivation of our conductance expression below {[}see Ref.\citep{Zarbo-2022}
and Appendix B{]}. In the case of a grounded TSC, symmetric QI-lead
couplings $(\Gamma)$ independent of $\mu_{\text{{Source}}}-\mu_{\text{{Drain}}}=\text{{eV}},$
being $\mu_{\text{{Source}}}=-\mu_{\text{{Drain}}}=\text{{eV}}/2,$
where $e$ is the elemental charge and V the corresponding bias-voltage,
the crossed Andreev reflection is suppressed and the conductance can
be split into\citep{Zarbo-2022} }
\begin{equation}
\mathcal{G}_{\text{{Total}}}=\mathcal{G}_{\text{{ET}}}+\mathcal{G}_{\text{{LAR}}},\label{eq:Gtotal}
\end{equation}
where ET and LAR stand for the electron tunneling and local Andreev
reflection processes, respectively, with
\begin{eqnarray}
\mathcal{G}_{\text{{ET}}}(\text{{eV}})=\frac{e^{2}}{2h}[\tau_{\alpha\bar{\alpha}}^{\text{{ET}}}(\text{{eV}}/2)+\tau_{\alpha\bar{\alpha}}^{\text{{ET}}}(-\text{{eV}}/2)],\label{eq:ET}
\end{eqnarray}
wherein $\alpha=\text{{Source}},$ $\bar{\alpha}=\text{{Drain} }$
and vice-versa, together with
\begin{eqnarray}
\mathcal{G}_{\text{{LAR}}}(\text{{eV}})=\frac{e^{2}}{2h}[\tau_{\alpha\alpha}^{\text{{LAR}}}(\text{{eV}}/2)+\tau_{\alpha\alpha}^{\text{{LAR}}}(-\text{{eV}}/2)],\label{eq:LAR}
\end{eqnarray}
in which the transmittances $\tau_{\alpha\bar{\alpha}}^{\text{{ET}}}=(2S+1)\Gamma^{2}|\langle\langle d_{\uparrow};d_{\uparrow}^{\dagger}\rangle\rangle_{\omega}|^{2}$
and $\tau_{\alpha\alpha}^{\text{{LAR}}}=(2S+1)\Gamma^{2}|\langle\langle d_{\uparrow}^{\dagger};d_{\uparrow}^{\dagger}\rangle\rangle_{\omega}|^{2}$
{are expressed in terms of the frequency-dependent
Green's functions (GFs) of type $\langle\langle A_{\sigma};B_{\sigma}\rangle\rangle_{\omega}=\sum_{m}\langle\langle A_{\sigma}|m\rangle\langle m|;B_{\sigma}\rangle\rangle_{\omega}$
(details in the Appendix A), due to the presence of the large spin,
in which the thermal average $\langle|m\rangle\langle m|\rangle=\frac{1}{2S+1}$
should be taken into account. Particularly, these GFs, are indeed
the time Fourier transform (with $k_{B}=\hbar=1)$ of}

{{}
\begin{equation}
\langle\langle A_{\sigma};B_{\sigma}\rangle\rangle_{\omega}=\int\sum_{m}\langle\langle A_{\sigma}(t)|m(t)\rangle\langle m(t)|;B_{\sigma}(0)\rangle\rangle e^{i\omega^{+}t}dt,\label{eq:FTofGF}
\end{equation}
where $\omega^{+}=\omega+\text{{i}}\eta^{+},$ with $\eta^{+}\rightarrow0$
and}

{{}
\begin{align}
\langle\langle A_{\sigma}(t)|m(t)\rangle\langle m(t)|;B_{\sigma}(0)\rangle\rangle & =-i\theta(t)\mathcal{Z}^{-1}\text{{Tr}}\{e^{-\mathcal{H}/T}\label{eq:TDGF}\\
 & [A_{\sigma}(t)|m(t)\rangle\langle m(t)|,B_{\sigma}(0)]_{+}\}\nonumber
\end{align}
stands for the time-dependent GF following double-brackets Zubarev
notation\cite{Zubarev-1960}, wherein $\text{{Tr}}$ gives the trace
over Eq.(\ref{eq:Htotal}) eigenstates, $\mathcal{Z}$ is the partition
function and $[...,...]_{+}$ as the anticommutator.}

{In practice, the GFs should be determined via the
standard equation-of-motion (EOM) approach{\citep{Flensberg-book}},
which in frequency domain, can be summarized as follows
\begin{equation}
\omega^{+}\langle\langle A_{\sigma};B_{\sigma}\rangle\rangle_{\omega}=\langle[A_{\sigma},B_{\sigma}]_{+}\rangle+\langle\langle[A_{\sigma},\mathcal{H}];B_{\sigma}\rangle\rangle_{\omega}.\label{eq:EOM}
\end{equation}
}Additionally, Ref.{\citep{Zarbo-2022}} also ensures that the QI
normalized DOS obeys the decomposition
\begin{equation}
\text{{DOS}}(\uparrow)=-\Gamma\text{{Im}}\langle\langle d_{\uparrow};d_{\uparrow}^{\dagger}\rangle\rangle_{\omega}=\tau_{\alpha\bar{\alpha}}^{\text{{ET}}}+\tau_{\alpha\alpha}^{\text{{LAR}}}.\label{eq:DOSelectron}
\end{equation}
As Eq.(\ref{eq:DOSelectron}) is bounded to unity, it describes the
electronic overall transmittance through the QI decomposed into ET
and LAR processes. Specially when it attains to its maximum value
at zero energy, i.e., the $\text{{DOS}}(\uparrow)(0)=1$ value gives
the electronic zero-frequency spectral weight. In this case, the regular
fermionic state of the QI is then made equally by the MZMs $\gamma_{A}$
and $\gamma_{B}.$ Equivalently, it means that the corresponding normalized
DOSs for such quasiparticles localize Majorana states with the same
spectral weights and as a result, the QI state is fully built by a
pair of resonant MZMs. It gives rise to the conductance $\mathcal{G}_{\text{{Total}}}(0)=\frac{e^{2}}{h}.$
Interestingly enough for $\text{{DOS}}(\uparrow)(0)=1/2,$ an isolated
ordinary MZM is found at the QI site and the zero-bias conductance
is characterized by the hallmark $\mathcal{G}_{\text{{Total}}}(0)=\frac{e^{2}}{2h}${\citep{Baranger-2011,Leakage-2014}}.
Such a case corresponds to an ideal infinite superconducting wire.{{}
However, there is a regime in which the value $\text{{DOS}}(\uparrow)(0)=1/2$
is still present for a finite wire }and due to the Ising interaction
between the large spin and the QI, the observation of the conductance
$\mathcal{G}_{\text{{Total}}}(0)=\frac{e^{2}}{2h}$ is ensured. This
emerges from the novel excitation that we introduce as the MIQ, in
particular, by driving the system into the sweet spot for the exchange
interaction $J$, namely, $J=J_{\text{{h}}}.$ In the latter, the
index ``h'' stands for the ``half-fermion'' special condition
of a MZM, which is produced by imposing $\text{{DOS}}(\uparrow)(0)=1/2,$
from where we extract $J_{\text{{h}}}$ for a given $S${[}see Fig.\ref{fig:Fig3}
b)-I{]}.

Therefore, in order to reveal the aforementioned Physics about the
system conductance, we should begin evaluating Eq.(\ref{eq:ET}) for
the electron tunneling process. Thus, the GF $\langle\langle d_{\uparrow};d_{\uparrow}^{\dagger}\rangle\rangle_{\omega}$
should be found via the EOM method, which gives
\begin{eqnarray}
\langle\langle d_{\uparrow};d_{\uparrow}^{\dagger}\rangle\rangle_{\omega} & = & \frac{1}{2S+1}\sum_{m}\frac{1}{\omega^{+}-\varepsilon_{\uparrow}-\frac{Jm}{2}+i\Gamma-\Sigma_{\text{{MFs}}}^{+m}},\nonumber \\
\label{eq:GFforET}
\end{eqnarray}
where $\Sigma_{\text{{MFs}}}^{+m}=K_{+}+(2t\Delta)^{2}K_{\text{{MFs}}}\Tilde{K}_{m}$
represents the self-energy correction due to the couplings of the
QI with the TSC and the large spin $S$. This also depends on the
following defined quantities
\begin{eqnarray}
K_{\text{{MFs}}}=\frac{\omega^{+}}{\omega^{2}-\varepsilon_{M}^{2}+2i\omega\eta^{+}-(\eta^{+})^{2}},\label{eq:SelfEMFs}
\end{eqnarray}
\begin{eqnarray}
K_{\pm}=\frac{\omega^{+}(\Delta^{2}+t^{2})\pm\varepsilon_{M}(t^{2}-\Delta^{2})}{\omega^{2}-\varepsilon_{M}^{2}+2i\omega\eta^{+}-(\eta^{+})^{2}}\label{eq:SelfEPlusMinus}
\end{eqnarray}
and
\begin{eqnarray}
\tilde{K}_{m}=\frac{K_{\text{{MFs}}}}{\omega^{+}+\varepsilon_{\uparrow}+\frac{Jm}{2}+i\Gamma-K_{-}}.\label{eq:SelfEmtil}
\end{eqnarray}
Concerning the LAR, the conductance of Eq.(\ref{eq:LAR}) needs the
evaluation of the anomalous GF $\langle\langle d_{\uparrow}^{\dagger};d_{\uparrow}^{\dagger}\rangle\rangle_{\omega}$
instead. After performing the EOM approach, it gives rise to
\begin{eqnarray}
\langle\langle d_{\uparrow}^{\dagger};d_{\uparrow}^{\dagger}\rangle\rangle_{\omega} & = & -\frac{1}{2S+1}\sum_{m}\frac{2t\Delta K_{m}}{\omega^{+}+\varepsilon_{\uparrow}+\frac{Jm}{2}+i\Gamma-\Sigma_{\text{{MFs}}}^{-m}},\nonumber \\
\label{eq:GFforLAR}
\end{eqnarray}
with $\Sigma_{MFs}^{-m}=K_{-}+(2t\Delta)^{2}K_{MFs}K_{m}$ and
\begin{eqnarray}
K_{m} & = & \frac{K_{\text{{MFs}}}}{\omega^{+}-\varepsilon_{\uparrow}-\frac{Jm}{2}+i\Gamma-K_{+}}.\label{eq:SelfEm}
\end{eqnarray}

However, if we want to know about the possibility of isolating MZMs
in the QI, the DOSs for $\gamma_{A}$ and $\gamma_{B}$ should be
found in order to examine the emergence of resonant states. To this
end, we invert Eq.(\ref{eq:Dup}) for $\gamma_{A}$ and $\gamma_{B},$
namely, $\gamma_{A}=(d_{\uparrow}^{\dagger}+d_{\uparrow})/\sqrt{2}$
and $\gamma_{B}=i(d_{\uparrow}^{\dagger}-d_{\uparrow})/\sqrt{2},$
{and we calculate the GFs} $\langle\langle\gamma_{A};\gamma_{A}\rangle\rangle_{\omega}$
and $\langle\langle\gamma_{B};\gamma_{B}\rangle\rangle_{\omega}.$
Consequently,
\begin{eqnarray}
\langle\langle\gamma_{j};\gamma_{j}\rangle\rangle_{\omega} & = & \frac{1}{2}[\langle\langle d_{\uparrow};d_{\uparrow}^{\dagger}\rangle\rangle_{\omega}+\langle\langle d_{\uparrow}^{\dagger};d_{\uparrow}\rangle\rangle_{\omega}\nonumber \\
 & + & \epsilon(\langle\langle d_{\uparrow}^{\dagger};d_{\uparrow}^{\dagger}\rangle\rangle_{\omega}+\langle\langle d_{\uparrow};d_{\uparrow}\rangle\rangle_{\omega})],\label{eq:GamajGamaj}
\end{eqnarray}
where $j=(A,B)$ corresponds to $\epsilon=(+1,-1),$ respectively.
Physically speaking, the sign reversal in $\epsilon$ can lead to
distinct quantum interference phenomena, in particular between those
encoded by the normal GFs ($\langle\langle d_{\uparrow};d_{\uparrow}^{\dagger}\rangle\rangle_{\omega}$
and $\langle\langle d_{\uparrow}^{\dagger};d_{\uparrow}\rangle\rangle_{\omega}$)
and the corresponding superconducting ($\langle\langle d_{\uparrow}^{\dagger};d_{\uparrow}^{\dagger}\rangle\rangle_{\omega}$
and $\langle\langle d_{\uparrow};d_{\uparrow}\rangle\rangle_{\omega}$).
To reveal such interference processes, we need just to find the GFs
$\langle\langle d_{\uparrow}^{\dagger};d_{\uparrow}\rangle\rangle_{\omega}$
and $\langle\langle d_{\uparrow};d_{\uparrow}\rangle\rangle_{\omega}$
to close the evaluation of $\langle\langle\gamma_{A};\gamma_{A}\rangle\rangle_{\omega}$
and $\langle\langle\gamma_{B};\gamma_{B}\rangle\rangle_{\omega}.$
By applying the EOM method, we conclude that
\begin{align}
\langle\langle d_{\uparrow}^{\dagger};d_{\uparrow}\rangle\rangle_{\omega} & =\frac{1}{2S+1}\sum_{m}\frac{1}{\omega^{+}+\varepsilon_{\uparrow}+\frac{Jm}{2}+i\Gamma-\Sigma_{\text{{MFs}}}^{-m}}\nonumber \\
\label{eq:GFdplusd}
\end{align}
and
\begin{align}
\langle\langle d_{\uparrow};d_{\uparrow}\rangle\rangle_{\omega} & =-\frac{1}{2S+1}\sum_{m}\frac{2\Delta t\tilde{K}_{m}}{\omega^{+}-\varepsilon_{\uparrow}-\frac{Jm}{2}+i\Gamma-\Sigma_{\text{{MFs}}}^{+m}}.\nonumber \\
\label{eq:GFdd}
\end{align}
Naturally, we define the normalized DOSs for $\gamma_{A}$ and $\gamma_{B}$
such as
\begin{equation}
\text{DOS}(\uparrow)[\gamma_{j}]=-\Gamma\text{{Im}}\langle\langle\gamma_{j};\gamma_{j}\rangle\rangle_{\omega}.\label{eq:DOSMajorana}
\end{equation}
This formula elucidates that when the quantity $\text{DOS}(\uparrow)[\gamma_{j}](0)=-\Gamma\text{{Im}}\langle\langle\gamma_{j};\gamma_{j}\rangle\rangle_{\omega}(0)=1$
is fulfilled, it can be recognized as the maximum Majorana quasiparticle
transmittance or its corresponding zero-frequency spectral weight.

Henceforward, we focus the attention on the case $\varepsilon_{\uparrow}=0$
(grounded SC). We perceive by inspecting Eqs.(\ref{eq:GFforET}) and
(\ref{eq:GFdplusd}), that Eq.(\ref{eq:DOSelectron}) becomes also
$\text{{DOS}}(\uparrow)=-\Gamma\text{{Im}}\langle\langle d_{\uparrow}^{\dagger};d_{\uparrow}\rangle\rangle_{\omega}.$
Additionally, $-\Gamma\text{{Im}}\langle\langle d_{\uparrow}^{\dagger};d_{\uparrow}^{\dagger}\rangle\rangle_{\omega}=-\Gamma\text{{Im}}\langle\langle d_{\uparrow};d_{\uparrow}\rangle\rangle_{\omega}.$
This in combination with Eqs.(\ref{eq:GamajGamaj}) and (\ref{eq:DOSMajorana})
allow us to establish that
\begin{equation}
\text{{DOS}}(\uparrow)=\frac{1}{2}(\text{DOS}(\uparrow)[\gamma_{A}]+\text{DOS}(\uparrow)[\gamma_{B}]).\label{eq:NewDOSelectron}
\end{equation}
Consequently, by taking into account this finding together with Eqs.(\ref{eq:Gtotal}),
(\ref{eq:ET}), (\ref{eq:LAR}) and (\ref{eq:DOSelectron}), we conclude
the providential equality as follows
\begin{equation}
\mathcal{G}_{\text{{Total}}}=\mathcal{G}_{\text{\ensuremath{\gamma_{A}}}}+\mathcal{G}_{\text{\ensuremath{\gamma_{B}}}},\label{eq:NewGtotal}
\end{equation}
where
\begin{align}
\mathcal{G}_{\text{\ensuremath{\gamma_{j}}}}(\text{{eV}}) & =\frac{e^{2}}{4h}[\text{DOS}(\uparrow)[\gamma_{j}](\text{{eV}}/2)+\text{DOS}(\uparrow)[\gamma_{j}](-\text{{eV}}/2)]\nonumber \\
\label{eq:GMajorana}
\end{align}
stands for the conductance contribution arising from the quasiparticle
$\gamma_{j}$ within the QI.

We highlight that Eq.(\ref{eq:NewGtotal}) introduces an alternative
perspective concerning the underlying Physics of the conductance in
Eq.(\ref{eq:Gtotal}): the ET and LAR quantum transport mechanisms
are revealed as the net effect of two Majorana quasiparticle conductances,
namely, the corresponding contributions arising from $\gamma_{A}$
and $\gamma_{B},$ respectively.{{}}{{}
In this context, our main findings hold for the constraint $J=J_{\text{{h}}}$
fulfilled}, thus characterizing the system sweet spot to produce the
MIQ. This regime consists of the maximum Majorana quasiparticle transmittance
$\text{DOS}(\uparrow)[\gamma_{j}](0)=1,$ surprisingly, fractionalized
and split into $\text{DOS}(\uparrow)[\gamma_{A}](0)=1/2$ and $\text{DOS}(\uparrow)[\gamma_{B}](0)=1/2.$
Despite such equipartition, the electronic transmittance is still
given by $\text{{DOS}}(\uparrow)(0)=1/2$ {[}see Fig.\ref{fig:Fig3}
b)-I{]} and according to Eq.(\ref{eq:GMajorana}), it ensures $\mathcal{G}_{\text{\ensuremath{\gamma_{A}}}}(0)=\frac{e^{2}}{4h}$
and $\mathcal{G}_{\text{\ensuremath{\gamma_{B}}}}(0)=\frac{e^{2}}{4h}.${{}
}{However, counterintuitively, solely $\mathcal{G}_{\text{\ensuremath{\gamma_{B}}}}(0)$
contains a MZM in the common sense}{,} i.e., a resonant
state, while $\mathcal{G}_{\text{\ensuremath{\gamma_{A}}}}(0)$ shows
a dip instead, but with the same magnitude of the peak in $\text{\ensuremath{\gamma_{B}}}.$
This is the reason why we call the contribution $\mathcal{G}_{\text{\ensuremath{\gamma_{B}}}}(0)=\frac{e^{2}}{4h}$
by $\mathcal{G}_{\text{{MIQ}}}(0),$ in attention to the emergent
MIQ. This is the unique MZM-type resonant state that appears in the
system, due to the interplay between the topological superconductivity
and the Ising Hamiltonian. In this case, Eqs.(\ref{eq:Gtotal}) and
(\ref{eq:NewGtotal}) ensure that when the MIQ emerges, $\mathcal{G}_{\text{\ensuremath{\gamma_{B}}}}(0)$
exhibits a maximum and $\mathcal{G}_{\text{\ensuremath{\gamma_{A}}}}(0)$
shows a minimum, in such a way that only $\mathcal{G}_{\text{{ET}}}(0)$
enters into $\mathcal{G}_{\text{{Total}}}(0)=\frac{e^{2}}{2h}.$ It
means that the LAR process is found entirely suppressed within this
regime. {The complete analysis here summarized will
be discussed in the next section.}

\section{Results }

{In the entire numerical analysis of Sec.A, we keep
constant $\varepsilon_{\uparrow}=0$ (grounded SC), $\lambda_{1}=2.12\Gamma$
($\Delta=t$) and perform variations in the parameters $\varepsilon_{M},S$
and $J.$ Partially for Sec.B, we assume $|\lambda_{1}|\sim|\lambda_{2}|$
{[}$\Delta\neq0,t\rightarrow0${]} from Eq.(\ref{eq:Htotal}) {[}Eq.(\ref{eq:HMZMs}){]}
to analyze the ABSs regime within the framework of the effective model\cite{Nonlocality_Majorana}.
We should remember that $\mu_{\text{{Source}}}-\mu_{\text{{Drain}}}=\text{{eV}}$
represents the bias- energy between source and drain leads, with the
choice $\mu_{\text{{Source}}}=-\mu_{\text{{Drain}}}=\text{{eV}}/2$
in our transport calculations. }

\subsection{{Majorana-Ising-type quasiparticle (MIQ) excitation}}

\begin{figure}[!]
\centering\includegraphics[width=1\columnwidth]{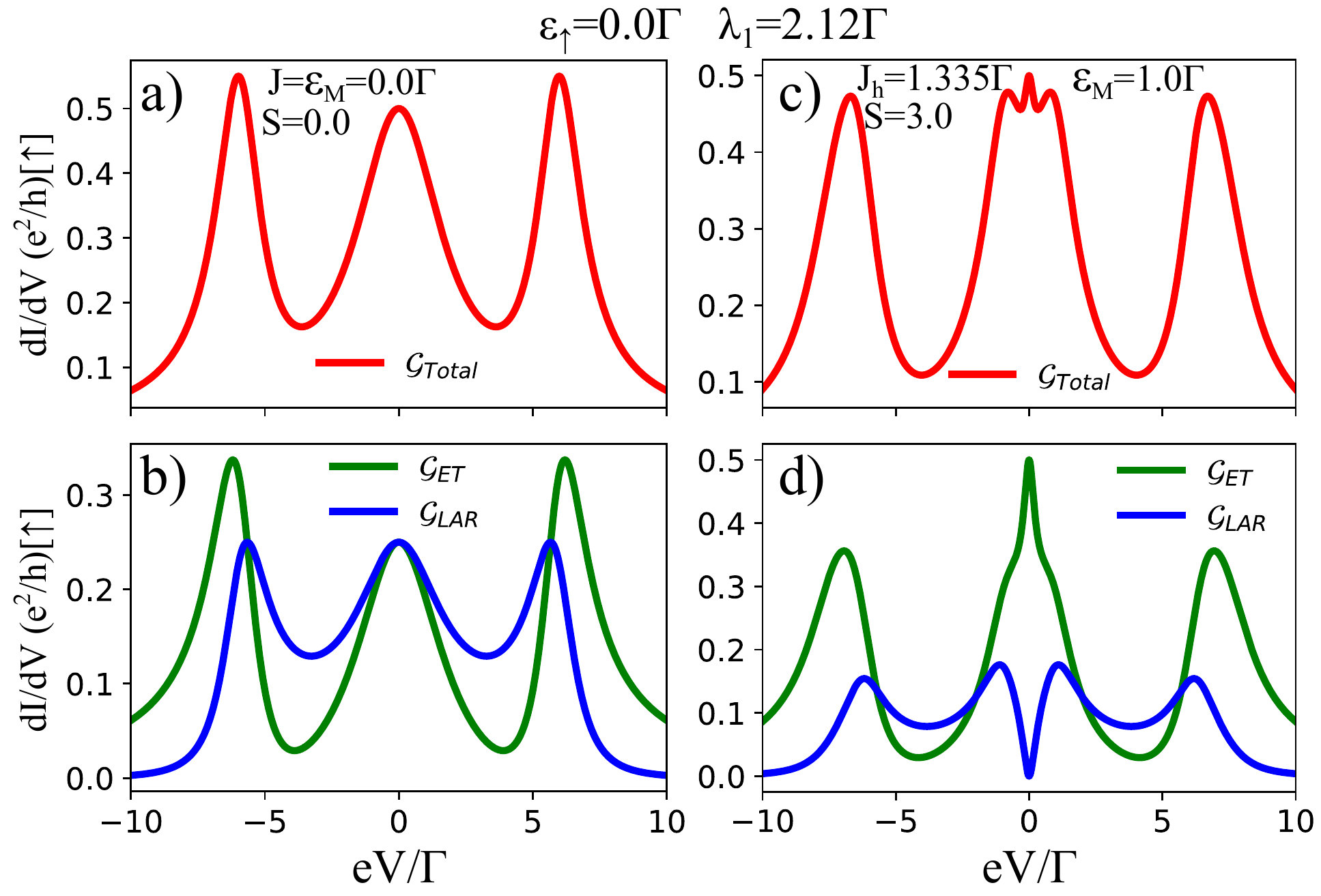} \caption{\label{fig:Fig2} (Color online) (a) Total differential conductance
$\mathcal{G}_{\text{{Total}}}$ {[}Eq.(\ref{eq:Gtotal}){]} versus
the source-drain bias-voltage $\text{{eV}}$ in units of the QI-lead
coupling $\Gamma$ with isolated MZMs $\gamma_{1}$ and $\gamma_{2}$
$(\varepsilon_{M}=0)$ in the absence of a spin-polarized tip $(S=0).$
As aftermath, a zero-bias peak (ZBP) emerges with amplitude $\mathcal{G}_{\text{{Total}}}(0)=\frac{e^{2}}{2h}$
and characterizes the leaking of the MZM $\gamma_{1}$ from the TSC
wire edge {[}Fig.\ref{fig:Fig1}(c){]} into the QI as $\gamma_{A}$
{[}inset a)-I of Fig.\ref{fig:Fig3}(a) and Refs.{\citep{Baranger-2011,Leakage-2014}}{]}.
(b) The ZBP conductance can be split into finite contributions from
the electron tunneling (ET) and local Andreev reflection (LAR) processes
{[}Eq.(\ref{eq:Gtotal}){]}, namely, $\mathcal{G}_{\text{{ET}}}(0)$
and $\mathcal{G}_{\text{{LAR}}}(0),$ respectively. This corresponds
to the ideal case of an infinite TSC wire, wherein these processes
compete on equal footing. (c) In the presence of a tip with $S=3$
and exchange coupling at the sweet spot $J=J_{\text{{h}}}=1.335\Gamma$
{[}inset b)-I of Fig.\ref{fig:Fig3}(b){]} for a finite wire $(\varepsilon_{M}=1\Gamma),$
the amplitude $\mathcal{G}_{\text{{Total}}}(0)=\frac{e^{2}}{2h}$
is still observed and denotes the existence of a MIQ $\gamma_{\text{{MIQ}}}$
within the QI, but with related differential conductance $\mathcal{G}_{\text{{MIQ}}}(0)=\frac{e^{2}}{4h}$
{[}Fig.\ref{fig:Fig3}(c) and Eqs.(\ref{eq:NewGtotal})-(\ref{eq:GMajorana}){]}.
(d) In the regime of (c), $\mathcal{G}_{\text{{LAR}}}(0)$ is completely
quenched and solely the term $\mathcal{G}_{\text{{ET}}}(0)$ contributes
to the ZBP.}
\end{figure}

\begin{figure}[!]
\centering\includegraphics[width=1\columnwidth]{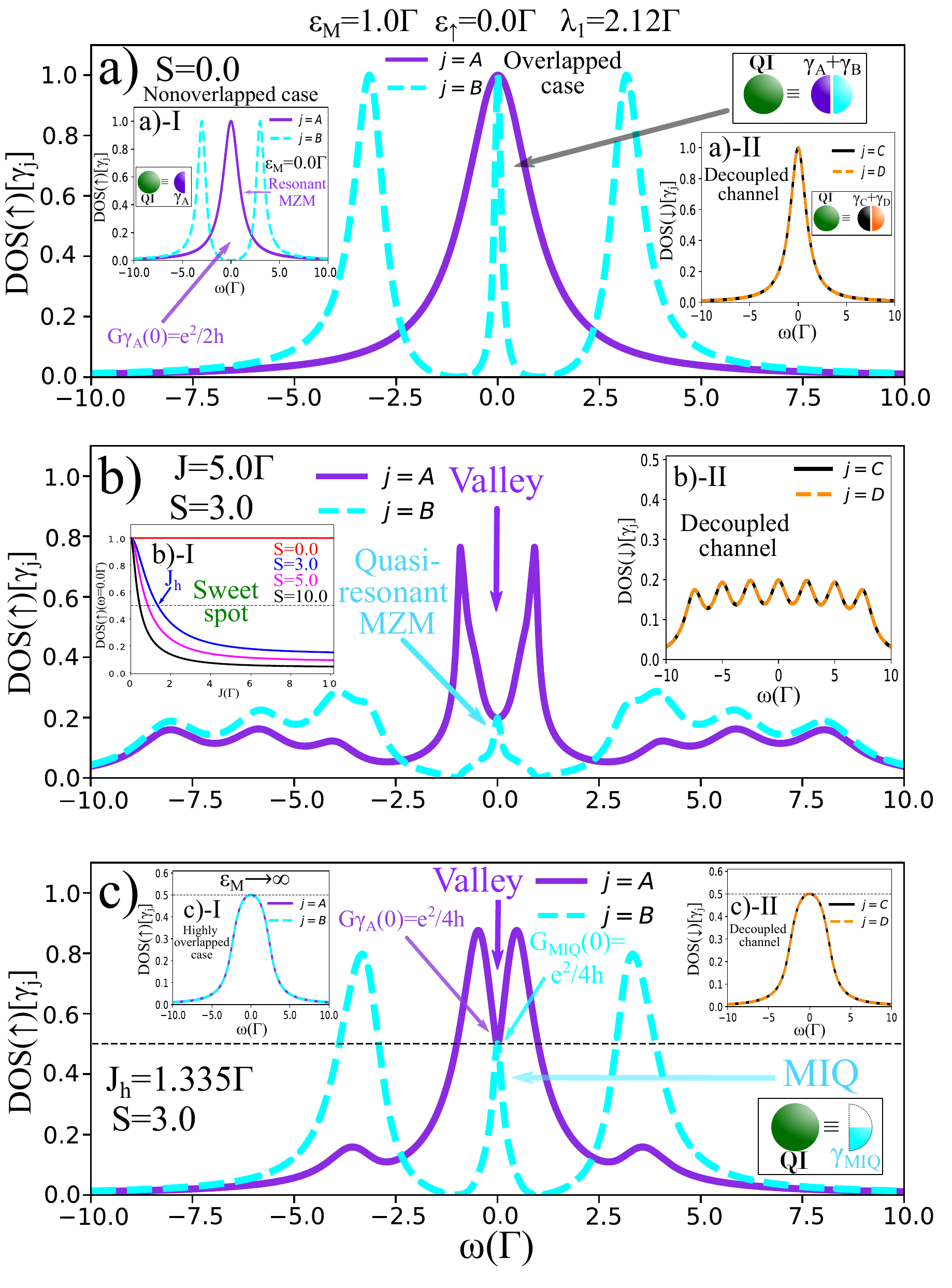} \caption{\label{fig:Fig3} (Color online) (a) The central panel shows the frequency
dependence of the normalized densities of states (DOSs) for the MZMs
$\gamma_{A}$ and $\gamma_{B}$ {[}Eq.(\ref{eq:DOSMajorana}){]} for
the spin-up channel within the QI by assuming the spin-polarized tip
absent and the MZMs $\gamma_{1}$ and $\gamma_{2}$ overlapped $(\varepsilon_{M}=1\Gamma)$
in the TSC. In each DOS we find a resonant MZM, which corresponds
to a ``half-fermionic'' state of the QI depicted as the green sphere
in the upper-right inset. This sphere has inner MZMs represented by
calottes to denote the half-fermion nature of the MZM. In the nonoverlapped
situation, only the calotte $\gamma_{A}$ prevails {[}inset a)-I{]}.
It leads to $\mathcal{G}_{\text{{Total}}}(0)=\mathcal{G}_{\gamma_{A}}(0)=\frac{e^{2}}{2h}$
in Fig.\ref{fig:Fig2}(a) {[}Eqs.(\ref{eq:NewGtotal})-(\ref{eq:GMajorana}){]}.
For the spin-down channel, which is decoupled from the TSC, the MZMs
$\gamma_{C}$ and $\gamma_{D}$ within the QI stay paired permanently
{[}inset a)-II{]}. (b) In the presence of $S=3$ and $J=5\Gamma$
for $\varepsilon_{M}=1\Gamma$ the DOS for $\gamma_{A}$ changes drastically
exhibiting a valley at $\omega=0$ instead of a peak, while for $\gamma_{B}$
the zero-mode is not-well defined. To restore the ZBP $\mathcal{G}_{\text{{Total}}}(0)=\frac{e^{2}}{2h}$
we should choose the right $J$ for a given $S,$ i.e., the sweet
spot $J=J_{\text{{h}}}$ {[}panel b)-I for the QI DOS of Eq.(\ref{eq:NewDOSelectron})
imposing $\text{{DOS}}(\uparrow)(0)=1/2${]}. In the spin-down channel,
the $\gamma_{C}$ and $\gamma_{D}$ DOSs are degenerate with a $2S+1$
multi-level structure {[}panel b)-II{]}. (c) The choice $J=J_{\text{{h}}}=1.335\Gamma$
for $S=3$ completely defines the valley and peak for the DOSs $\gamma_{A}$
and $\gamma_{B},$ respectively, where in the latter we introduce
$\gamma_{B}\equiv\gamma_{\text{{MIQ}}}$ as the Majorana-Ising quasiparticle,
once the peak clearly represents the unique MZM isolated in the system.
It is characterized by a ZBP given by $\mathcal{G}_{\text{{MIQ}}}(0)=\frac{e^{2}}{4h}$
pictorially illustrated by the half-calotte in the lower-right inset
for the green sphere representing the QI fermionic state. However,
$\mathcal{G}_{\text{{Total}}}(0)=\mathcal{G}_{\gamma_{A}}(0)+\mathcal{G}_{\text{{MIQ}}}(0)=\frac{e^{2}}{2h}.$ }
\end{figure}

In Fig.\ref{fig:Fig2} we present, for the QI of Fig.\ref{fig:Fig1},
the total conductance of Eq.(\ref{eq:Gtotal}) as a function of the
bias-voltage $\text{{eV}}/\Gamma.$ Particularly in Fig.\ref{fig:Fig2}(a)
the ideal case is considered, i.e., the TSC wire is perfectly infinite
$(\varepsilon_{M}=0)$ and the large spin is found turned-off $(S=0).$
This case is well-known, being characterized by the ZBP in the conductance
given by $\mathcal{G}_{\text{{Total}}}(0)=\frac{e^{2}}{2h}${\citep{Baranger-2011,Leakage-2014}}.
Interestingly enough, this ZBP in the conductance represents the isolated
MZM $\gamma_{1}$ originally attached to one edge of the TSC wire,
which leaks towards the QI site in the form of the MZM $\gamma_{A}.$
The MZM leakage from the TSC edge into the QI is then characterized
by the $\text{DOS}(\uparrow)[\gamma_{A}](0)=1$ and $\text{DOS}(\uparrow)[\gamma_{B}](0)=0.$
{We will provide extra details concerning this issue
in the discussion of the inset a)-I of Fig.\ref{fig:Fig3}(a)}{.}
In the other hand, the satellite peaks in Fig.\ref{fig:Fig2}(a) are
the aftermath of the splitting arising from the condition $i\sqrt{2}\lambda_{1}\gamma_{B}\gamma_{1}$
given by Eq.(\ref{eq:HMZMs}){{} for the }\textit{{poor
man's Majorana}}{s regime of the system}\cite{Flensberg_2012(Poor),Kouwenhoven_2023}.
Additionally, we have made explicit via Eq.(\ref{eq:Gtotal}) that
the ET and LAR processes compose the total conductance. Thus, such
a feature can be viewed in Fig.\ref{fig:Fig2}(b), where we notice,
in particular, for the ZBP conductance that, the ET and LAR split
equally. Here we propose that it is still achievable to obtain $\mathcal{G}_{\text{{Total}}}(0)=\frac{e^{2}}{2h}$
for a finite TSC wire and to perform also the isolation of a Majorana
quasiparticle at the QI site. In our setup, such an excitation rises,
in particular, dressed by the Ising interaction. To this end, an integer
large spin $S$ should be accounted for and be coupled to the QI with
a special value in the exchange interaction $J.$ Thus, by evaluating
$J=J_{\text{{h}}},$ the amplitude $\mathcal{G}_{\text{{Total}}}(0)=\frac{e^{2}}{2h}$
{[}Eq.(\ref{eq:NewGtotal}){]} finally becomes restored. However,
we will verify that such a configuration corresponds to isolate a
peculiar MZM, namely $\gamma_{B},$ with a resonant peak characterized
by the spectral weight $\text{DOS}(\uparrow)[\gamma_{B}](0)=1/2$
{[}Eq.(\ref{eq:DOSMajorana}){]}, while for $\gamma_{A}$ we have
the same amplitude, i.e., $\text{DOS}(\uparrow)[\gamma_{A}](0)=1/2,$
but with a dip instead.

Now, we consider the presence of a large spin $S.$ Figs.\ref{fig:Fig2}(c)
and (d) show the total conductance in the presence of $S=3$ and $J_{\text{{h}}}=1.335\Gamma$
{[}see inset b)-I of Fig.\ref{fig:Fig3}(b){]} for a finite TSC with
$\varepsilon_{M}=\Gamma.$ The ZBP conductance in Fig.\ref{fig:Fig2}(c)
is $\mathcal{G}_{\text{{Total}}}(0)=\frac{e^{2}}{2h}$ as expected,
but the decomposition into ET and LAR channels described in Fig.\ref{fig:Fig2}(d)
reveals a striking result: solely ET survives, while LAR is completely
suppressed at zero-bias. Below we will verify that the LAR suppression
corresponds to a quasiparticle localization in the DOS for $\gamma_{B},$
which leads to $\text{DOS}(\uparrow)[\gamma_{B}](0)=1/2.$ Thus, according
to Eq.(\ref{eq:GMajorana}), such a finite DOS contributes to a conductance
$\mathcal{G}_{\text{\ensuremath{\gamma_{B}}}}(0)=\mathcal{G}_{\text{{MIQ}}}(0)=\frac{e^{2}}{4h},$
where we define the MIQ $\gamma_{B}\equiv\gamma_{\text{{MIQ}}}.$

In order to understand the emergence of the MIQ, we begin with the
trivial case in Fig.\ref{fig:Fig3}(a): the central panel discriminates
the electronic $\text{{DOS}}(\uparrow)$ into the corresponding DOSs
for $\gamma_{A}$ and $\gamma_{B}$ {[}Eq.(\ref{eq:DOSMajorana}){]}
with $S=0$ and $\varepsilon_{M}=\Gamma.$ This case is regarded as
trivial, once we verify that in both the DOSs $\gamma_{A}$ and $\gamma_{B}$
resonant states pinned at zero-energy $\omega=0$ are present. Schematically,
the QI fermionic state can be imagined as a sphere formed by two MZMs
depicted by two calottes. This is the manner we outline pictorially
the two zero-energy resonant states, the so-called MZMs in the DOSs
$\gamma_{A}$ and $\gamma_{B}.$ This sketch can be found in the upper-right
inset of Fig.\ref{fig:Fig3}(a), in which each calotte symbolizes
the ``half-fermionic'' character of the MZM. Equivalently, a calotte
occupies the half-volume of the sphere and as it corresponds to a
MZM, it can be surely characterized by $\text{DOS}(\uparrow)[\gamma_{j}](0)=1.$
As aftermath, according to Eqs.(\ref{eq:NewDOSelectron})-(\ref{eq:GMajorana}),
these two MZMs lead to the zero-bias conductance peak $\mathcal{G}_{\text{{Total}}}(0)=\frac{e^{2}}{h}.$
Concerning the satellite peaks in the $\text{DOS}(\uparrow)[\gamma_{B}],$
they occur due to the overlap between the MZMs $\gamma_{B}$ and $\gamma_{1}.$
In the other hand, the inset panel a)-II of Fig.\ref{fig:Fig3}(a)
shows the spin-down channel, which is the one decoupled from the TSC.
To analyze it on the same footing as the spin-up channel, {we
assume artificially $\varepsilon_{\downarrow}=0$} and verifies that
both the MZMs $\gamma_{C}$ and $\gamma_{D},$ which constitute this
spin sector of the QI, are then identified exactly by degenerate resonant
states. The evaluation of the DOSs for the spin-down sector just employs
the GFs for the spin-up sector, but it disregards the superconducting
terms. As none of these MZMs overlap with $\gamma_{1},$ a full superposition
of the lineshapes for these MZMs manifests in the profiles of the
DOSs $\gamma_{C}$ and $\gamma_{D}.$ Therefore, satellite peaks do
not emerge. Now, let us go back to discuss the spin-up channel. We
should pay particular attention to the case $\varepsilon_{M}=0$ depicted
in the inset panel a)-I of Fig.\ref{fig:Fig3}(a), which corresponds
to the nonoverlapped situation between $\gamma_{1}$ and $\gamma_{2}.$
This scenario is the ideal one and it contains the pillars for the
conductance behavior of Fig.\ref{fig:Fig2}(a). Notice that $\gamma_{B}$
overlaps with $\gamma_{1}$ leading to satellite peaks in $\text{DOS}(\uparrow)[\gamma_{B}]$
and $\text{DOS}(\uparrow)[\gamma_{B}](0)=0,$ while $\gamma_{A}$
is found isolated and localized as a well-defined resonant zero-mode
with spectral weight $\text{DOS}(\uparrow)[\gamma_{A}](0)=1.$ Thereafter,
$\mathcal{G}_{\text{{Total}}}(0)=\mathcal{G}_{\text{\ensuremath{\gamma_{A}}}}(0)=\frac{e^{2}}{2h}.$
This case is well-known in literature {\citep{Baranger-2011,Leakage-2014}}
and points out that the MZM $\gamma_{A}$ contributes to the conductance
as a resonant state in contrast to $\gamma_{B}$, which shows a gap
in $\text{DOS}(\uparrow)[\gamma_{B}]$ around zero-bias. This latter
prevents a finite conductance, i.e., $\mathcal{G}_{\text{\ensuremath{\gamma_{B}}}}(0)=0.$
Below, we will see that by turning-on the exchange $J$ for a given
$S,$ the spectral profiles for the $\text{DOS}(\uparrow)[\gamma_{A}]$
and $\text{DOS}(\uparrow)[\gamma_{B}]$ will exhibit a multi-level
structure. Additionally, these densities will be responsible, according
to Eq.(\ref{eq:GMajorana}), by a nonquantized $\mathcal{G}_{\text{{Total}}}(0)\neq\frac{e^{2}}{h}$
in Eq.(\ref{eq:NewGtotal}). In the situation of arbitrary $J,$ the
contributions to $\mathcal{G}_{\text{{Total}}}(0)\neq0$ will not
arise from well-defined resonant zero-mode states in $\text{DOS}(\uparrow)[\gamma_{A}]$
and $\text{DOS}(\uparrow)[\gamma_{B}].${{} The Majorana
fermion localization within the QI as a resonant zero-mode state will
only occur for the sweet spot $J=J_{\text{{h}}}.$}

\begin{figure*}[!]
\centering\includegraphics[width=2\columnwidth]{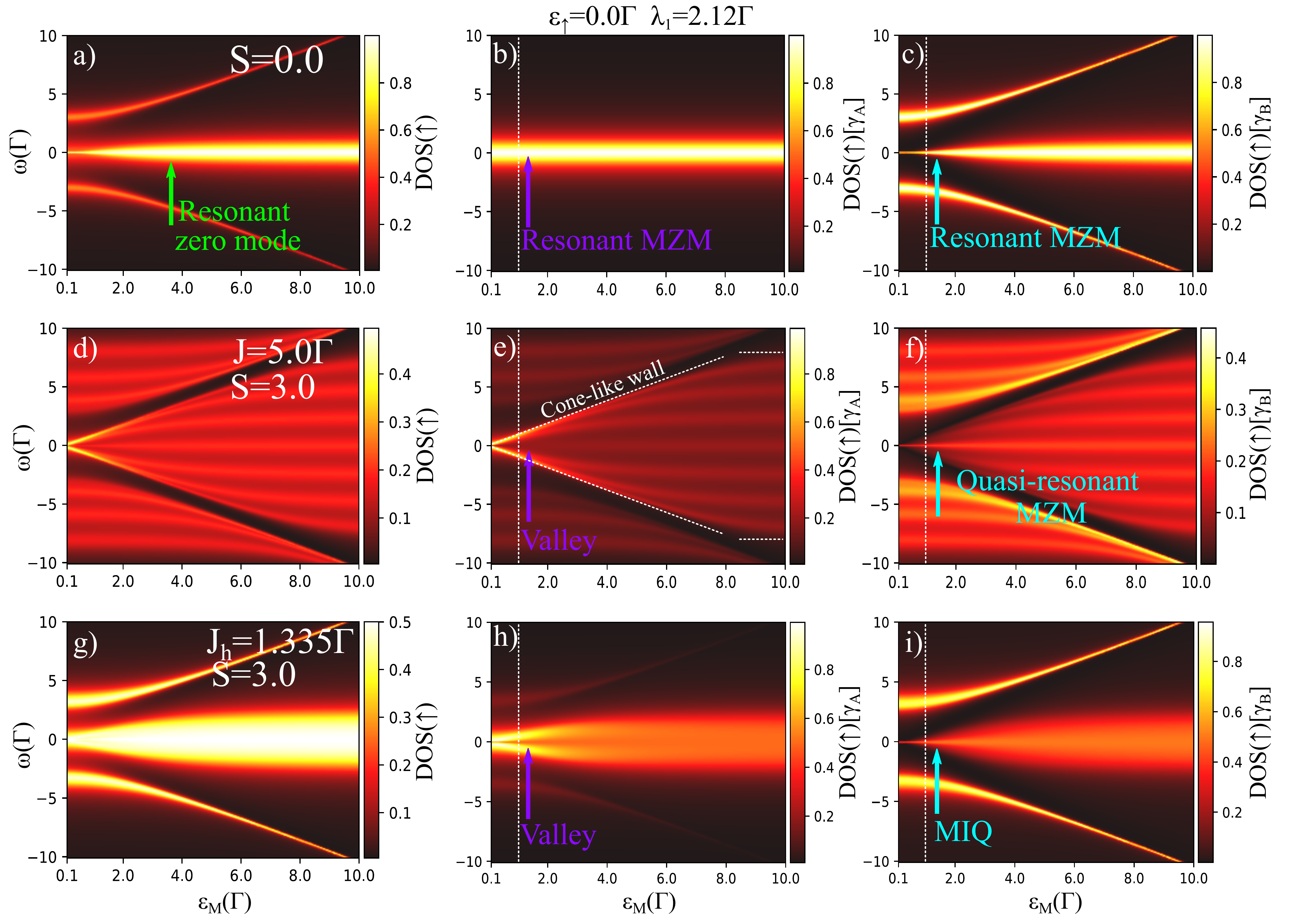} \caption{\label{fig:Fig4} (Color online) Color maps of Eqs.(\ref{eq:DOSelectron})
and (\ref{eq:DOSMajorana}) for the QI, $\gamma_{A}$ and $\gamma_{B}$
DOSs, respectively, spanned by the frequency $\omega$ and $\varepsilon_{M}$:
(a) In the case $S=0,$ the ZBP in the $\text{{DOS}}(\uparrow)$ arises
from the individual ZBPs found in $\text{{DOS}}(\uparrow)[\gamma_{A}]$
{[}panel(b){]} and $\text{{DOS}}(\uparrow)[\gamma_{B}]$ {[}panel(c){]},
due to the MZMs $\gamma_{A}$ and $\gamma_{B}$ that act as the building-blocks
of the QI state $\varepsilon_{\uparrow}=0$ {[}Eq.(\ref{eq:NewDOSelectron}){]}.
They reveal that the system does not contain isolated MZMs and that
the overlapped $\gamma_{1}$ and $\gamma_{2}$ split the zero-mode
energy state leading to the upper and lower arcs in panels (a) and
(c). Panels (d)-(f) reveal the influence of $S=3$ and $J=5\Gamma$
on the QI, where we can clearly see a $2S+1$ multi-level structure
centered at $\omega=0.$ The upper and lower arcs, distinctly, show
$2S$ levels each. The central region of (e) has a valley-type structure
in contrast to that for (f), which points out a quasi-resonant MZM.
In both the cases, the $2S+1$ multi-level structure is delimited
by cone-like walls up to a threshold in $\varepsilon_{M}$ (not indicated),
where above it the $\text{{DOS}}(\uparrow)[\gamma_{A}]$ and $\text{{DOS}}(\uparrow)[\gamma_{B}]$
exhibit a zero-mode with the lines of the walls parallel to each other.
For this situation, the TSC plays no role and the MZMs $\gamma_{A}$
and $\gamma_{B}$ stay paired as in (a)-(c). Panels (g)-(i) hold for
the sweet spot $J=J_{\text{{h}}}=1.335\Gamma,$ where we have the
zero-frequency valley and peak well-resolved in the $\text{{DOS}}(\uparrow)[\gamma_{A}]$
and $\text{{DOS}}(\uparrow)[\gamma_{B}],$ respectively. In the latter,
the MIQ rises as aftermath of the partial merge of the $2S+1$ multi-level
structure centered at $\omega=0$ upon decreasing the coupling $J.$
We call the attention that the vertical lines represent slice cuts
of the cases profoundly explored in Figs.\ref{fig:Fig2} and \ref{fig:Fig3}.}
\end{figure*}

Fig.\ref{fig:Fig3}(b) treats the presence of a large spin coupled
to the QI for $\varepsilon_{M}=\Gamma.$ As we can notice, the Ising
interaction $J=5\Gamma$ for $S=3$ clearly affects the spectral profiles
of $\text{DOS}(\uparrow)[\gamma_{A}]$ and $\text{DOS}(\uparrow)[\gamma_{B}]$.
Basically, it introduces a multi-level structure for the satellite
peaks and modifies drastically the MZM localization around the zero-bias.
Surprisingly, the $\text{DOS}(\uparrow)[\gamma_{A}]$ presents a valley
(dip) at zero-energy and a quasi-resonant MZM rises in $\text{DOS}(\uparrow)[\gamma_{B}].$
As a matter of fact, the latter cannot be considered a well-defined
resonant MZM: its spectral weight is not exactly an integer or semi-integer
number, and the lineshape broadening does not obey a \textit{{lorentzian-like}}{{}
form}. The lineshapes of such spectral densities are then distinct,
but counterintuitively, their zero-bias values coincide, i.e., $\text{DOS}(\uparrow)[\gamma_{A}](0)=\text{DOS}(\uparrow)[\gamma_{B}](0).$
As the spin-up sector of the QI is the one coupled to the TSC, the
spectral profiles in the $\text{DOS}(\uparrow)[\gamma_{A}]$ and $\text{DOS}(\uparrow)[\gamma_{B}]$
do not follow strictly the standard angular momentum theory for the
Zeeman splitting {[}Eq.(\ref{eq:HIsing}){]}. Usually, this theory
ensures that for an integer $S,$ $2S$ symmetrically displaced levels
around the corresponding at $\omega=\varepsilon_{\uparrow}=0$ should
emerge. Here, indeed we observe a mirror-symmetric set with $S$ energy
bands below and above the zero-bias, respectively, where nearby peculiar
spectral structures rise. We reveal that the profiles differ significantly
in this range, as aftermath of the nontrivial interplay between the
TSC and Ising exchange term. In Figs.\ref{fig:Fig4}(d)-(f) we will
see that the manifestation of this effect {lies within}
a region comprised by \textit{cone-like walls }spanned by $\omega$
and $\varepsilon_{M}$ in the DOSs of the system. Within the cone,
the Zeeman splitting becomes unusual and the energy spacing between
the levels is simultaneously governed by $J$ and $\varepsilon_{M}.$
Besides, solely the spin-down channel shows standard Zeeman splitting,
once it does not perceive the TSC. This can be verified in the inset
panel b)-II of Fig.\ref{fig:Fig3}(b), where the DOSs for $\gamma_{C}$
and $\gamma_{D},$ as expected, present the ordinary $2S+1$ multi-level
structure ensured by Eq.(\ref{eq:HIsing}). We highlight that upon
decreasing the exchange parameter $J$, the restoration of the conductance
$\mathcal{G}_{\text{{Total}}}(0)=\frac{e^{2}}{2h}$ can be still allowed.
Thus, we should remember that such a conductance arises from the fulfillment
of the condition $\text{{DOS}}(\uparrow)(0)=1/2.$ Particularly in
the inset panel b)-I of Fig.\ref{fig:Fig3}(b), we show exactly the
points where this happens by considering several values of $S$ and
$\varepsilon_{M}\neq0.$ Particularly for $S=3,$ this sweet spot
occurs for $J=J_{\text{{h}}}=1.335\Gamma$ and its dependence on $\varepsilon_{M}$
is revealed as very weak according to our numerical calculations (not
shown). It means that $J=J_{\text{{h}}}=1.335\Gamma$ still keeps
the value $\text{{DOS}}(\uparrow)(0)=1/2$ while $\varepsilon_{M}$
does not exceed very much $\Gamma$ ($\varepsilon_{M}\gg\Gamma$).
Experimentally speaking, $J$ can be tuned by changing the tip-QI
vertical distance. Thus for $\varepsilon_{M}\neq0,$ $\mathcal{G}_{\text{{Total}}}(0)$
drops from $\frac{e^{2}}{h}$ to $\frac{e^{2}}{2h}$ when $J=J_{\text{{h}}}.$
Hence, at the sweet spot, if one knows previously the spin $S$ of
the tip, $J_{\text{{h}}}$ can be extracted from Fig.\ref{fig:Fig3}
b)-I or vice-versa.

In Fig.\ref{fig:Fig3}(c) we present the case $J=J_{\text{{h}}}=1.335\Gamma$
that leads to our main finding: {a localization} of
a resonant state at zero-energy in $\text{DOS}(\uparrow)[\gamma_{B}],$
with spectral weight $\text{DOS}(\uparrow)[\gamma_{B}](0)=1/2.$ The
latter amplitude points out that the ordinary MZM now is fractionalized
and the value $\text{DOS}(\uparrow)[\gamma_{B}](0)=1$ is not present
anymore. This fractionalized MZM, the called MIQ by us, then leads
to a conductance $\mathcal{G}_{\text{\ensuremath{\gamma_{B}}}}(0)=\mathcal{G}_{\text{{MIQ}}}(0)=\frac{e^{2}}{4h}$
as ensured by Eq.(\ref{eq:GMajorana}). The value $\text{DOS}(\uparrow)[\gamma_{B}](0)=1/2$
can be pictorially viewed by the half-calotte found in the lower-right
inset of Fig.\ref{fig:Fig3}(c), which symbolizes the MZM fractionalization
within the QI state. Nevertheless, to complete the total conductance
$\mathcal{G}_{\text{{Total}}}(0)=\frac{e^{2}}{2h},$ the zero-bias
value for $\text{DOS}(\uparrow)[\gamma_{A}](0)$ should coincide,
i.e., $\text{DOS}(\uparrow)[\gamma_{A}](0)=1/2$ and as aftermath,
$\mathcal{G}_{\text{\ensuremath{\gamma_{A}}}}(0)=\frac{e^{2}}{4h}$
as well. Here we emphasize that in $\text{DOS}(\uparrow)[\gamma_{A}](0)$
a quantum destructive interference manifests and a resonant state
does not rise at zero-bias. Moreover, it is capital to clarify the
underlying mechanism to produce the MIQ: the key idea is the decrease
of $J,$ thus forcing the merge of the $2S$ side bands (satellite
peaks) of the system towards the zero-energy, where a resonant level
is pinned. As a result, this sum of amplitudes for the satellite peaks
interferes constructively at zero-energy giving rise to $\text{{DOS}}(\uparrow)(0)=1/2$
for $J=J_{\text{{h}}}.$ Further, our findings do not depend on the
sign of $J$ and in case of a semi-integer large spin, the multi-level
structure is {even} and a zero-energy is absent in
the spectrum. In this manner, the MIQ cannot be excited. Concerning
the inset panels c)-I and c)-II of Fig.\ref{fig:Fig3}(c), we verify
that for $\varepsilon_{M}\gg\Gamma$ (or $\varepsilon_{M}\rightarrow\infty$)
the DOSs for the Majorana fermions are degenerate. For extremely short
TSC wires ($\varepsilon_{M}\rightarrow\infty$), the energy level
$\varepsilon_{M}$ for the orbital $f$ {[}Eq.(\ref{eq:HMZMs}){]}
is highly off resonance the QI energy $\varepsilon_{\uparrow}=0,$
thus making the QI and the TSC to decouple from each other. Thus,
the spin-up channel behaves as the corresponding spin-down, which
is the one permanently decoupled from the TSC. This implies that the
MIQ cannot be seen for very short wires.

Fig.\ref{fig:Fig4} summarizes our findings exhibiting color maps
of Eqs.(\ref{eq:DOSelectron}) and (\ref{eq:DOSMajorana}) for the
electronic and Majorana DOSs, respectively and spanned by $\omega$
and $\varepsilon_{M}.$ Panels (a)-(c) describe the case $S=0,$ which
is characterized by $\text{{DOS}}(\uparrow)(0)=1$ and $\text{DOS}(\uparrow)[\gamma_{A}](0)=\text{DOS}(\uparrow)[\gamma_{B}](0)=1.$
This corresponds to the trivial regime where two MZMs localize around
zero-bias and appear at the QI site. Such a characteristic {lies
on} the zero-bias peaks and appears as horizontal lines in the representation
of panels (a)-(c) for any finite value of $\varepsilon_{M}.$ As we
can see, the satellite peaks in (a) arise from (c). By turning-on
the Ising interaction with $S=3$ and $J=5\Gamma,$ the spectral profiles
of the DOSs acquire distinct patterns: the satellite peaks obey approximately
the standard angular momentum theory for the Zeeman splitting, thus
exhibiting $2S$ split side-bands. In this situation, the linear dependence
on the exchange parameter $J$ is lacking. Besides, the central regions
of panels (a)-(c) are converted into the domains delimited by \textit{cone-like
walls} as those found in (d)-(f).\textit{ }While these walls persist
up to a threshold in $\varepsilon_{M}$ (not marked in the figure),
a sophisticated interplay between the TSC and the Ising interaction
rules the Physics of the system and allows the possibility of the
MIQ existence. Notice that for $\varepsilon_{M}>\Gamma,$ a $2S+1$
multi-level central structure finally becomes resolved. It is worth
mentioning that for $\varepsilon_{M}=\Gamma,$ the line cuts in Fig.\ref{fig:Fig4}
given by the vertical dashed lines, then correspond to the cases discussed
in detail in Fig.\ref{fig:Fig3}. {We would like to
mention that the choice $\varepsilon_{M}=\Gamma$ corresponds to a
strong limit, just to better resolve our findings. However, while
$\varepsilon_{M}$ stays within the aforementioned }\textit{{cone-like
walls,}}{{} the effect persists}. In panels (e) and
(f) we notice the rising of the valley and the quasi-resonant MZM
spectral structures, respectively upon increasing $\varepsilon_{M}.$
However, much above the threshold in $\varepsilon_{M}$, the linear
spacing in $J$ for the Zeeman splitting is restored and this situation
is that delimited by the marked horizontal dashed lines. Finally,
panels (g)-(i) show the merge of the multi-level structure in the
sweet spot $J=J_{\text{{h}}}=1.335\Gamma$ leading to the emergence
of the MIQ in the Majorana channel $\text{\ensuremath{\gamma_{B}}},$
while the valley continues in the channel $\text{\ensuremath{\gamma_{A}}}.$
Therefore, within \textit{the cone-like walls} domain, the sector
$\text{\ensuremath{\gamma_{B}}}$ of Majorana fermions for the QI
makes explicit a constructive interference process at zero-bias, while
the corresponding in $\text{\ensuremath{\gamma_{A}}}$ displays a
destructive behavior. In this regime, the conductance $\mathcal{G}_{\text{{LAR}}}(0)$
becomes fully quenched and just $\mathcal{G}_{\text{{ET}}}(0)$ contributes
to $\mathcal{G}_{\text{{Total}}}(0)=\frac{e^{2}}{2h}$ {[}Fig.\ref{fig:Fig2}(d){]}.

\subsection{{Poor man's Majoranas, parity qubit and ABSs regime}}

\begin{figure}
\centering\includegraphics[width=1\columnwidth]{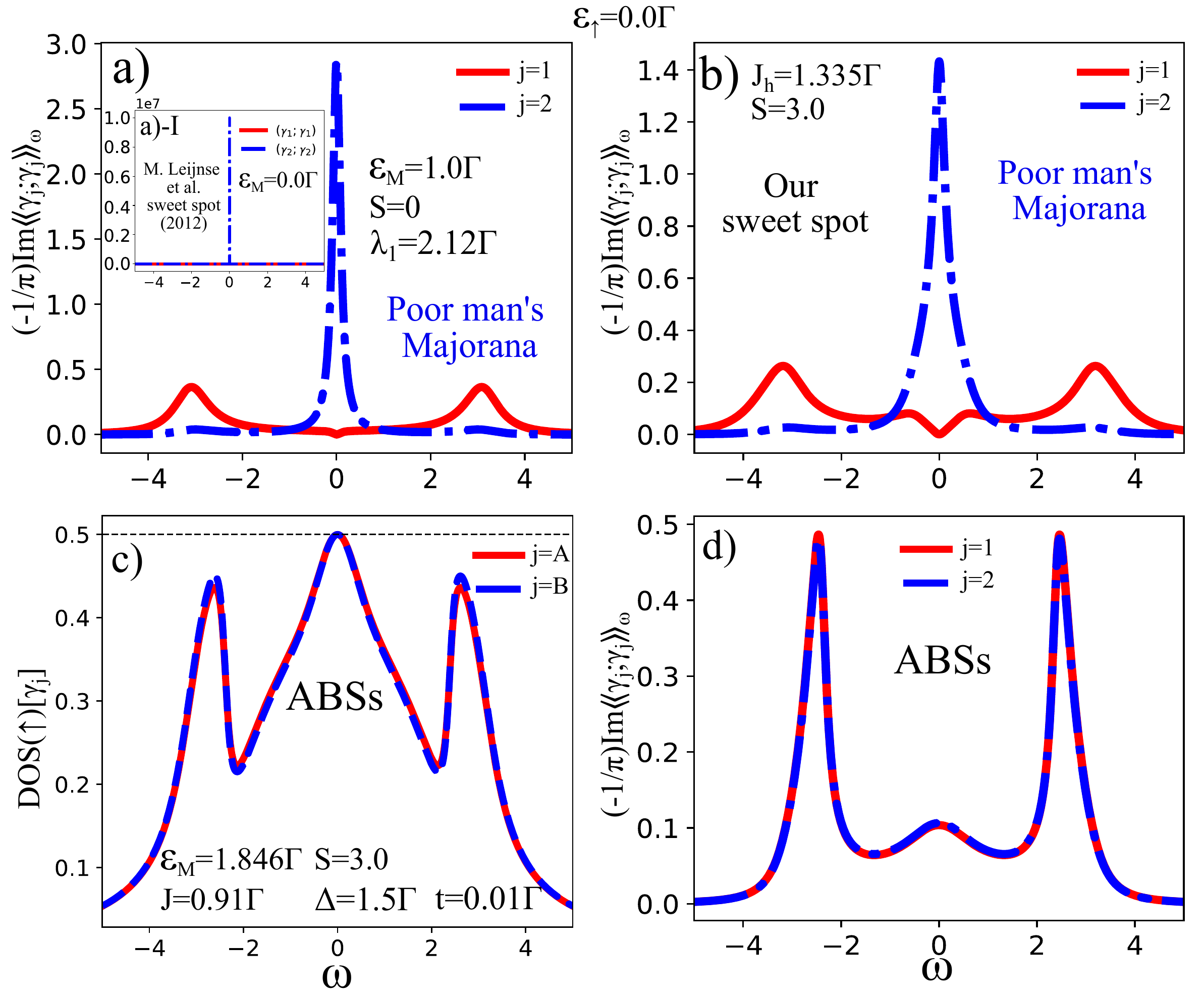} \caption{\label{fig:Fig5} {(Color online)} {Panels
(a) and (b) show M. Leijnse }\textit{{et al}}{\cite{Flensberg_2012(Poor)}
and our sweet spots, respectively for }\textit{{poor
man's Majorana}}{s. The Andreev bound states (ABSs)
regime of Refs.\cite{Nonlocality_Majorana,Ricco_2019} appear depicted
in panels (c) and (d). Additionally, fractionalization of zero-modes
also occurs in the latter.}}
\end{figure}

{In this section, we clarify that the MIQ excitation
can be classified as a}\textit{{{} poor man's Majorana
}}{\cite{Flensberg_2012(Poor),Kouwenhoven_2023} and
to demonstrate such, the analysis of the DOSs for the MZMs of Eq.(\ref{eq:HMZMs})
placed at the Kitaev dimer right edge is performed. To accomplish
this goal, we employ pertinent GFs from the Appendix C. Additionally,
we discuss the possibility of having a MIQ excitation-based parity
qubit for quantum computing purposes\cite{Flensberg_2012(Poor)} and
the ABSs regime within the effective model of Eq.(\ref{eq:Htotal})\cite{Nonlocality_Majorana}.
To this end, let us focus again on the left edge of the Kitaev dimer,
i.e., the QI. This system part description is found in central panel
of Fig.\ref{fig:Fig3}(a) ($\varepsilon_{M}=\Gamma$) and its inset
a)-I ($\varepsilon_{M}=0$), which account for the QI operator $d_{\uparrow},$
where the MZMs $\gamma_{A}$ and $\gamma_{B}$ reside {[}Eq.(\ref{eq:Dup}){]}.
Particularly, these panels show two resonant and single MZMs, respectively,
while in Fig.\ref{fig:Fig5}(a), a resonant MZM appears permanently
in the DOS $(-1/\pi)\text{{Im}}\langle\langle\gamma_{2};\gamma_{2}\rangle\rangle_{\omega}$
(not normalized, once leads are lacking at this side) for $\gamma_{2}$
in both the situations, thus describing the MZM placed at the TSC
right edge. Such a characteristic is entirely understood within the
theoretical framework for the Kitaev dimer described in Ref.\cite{Flensberg_2012(Poor)}.
The latter points out that systems based on Eq.(\ref{eq:HMZMs}) for
QIs tunnel and Andreev-coupled, which in our case are given by the
$d_{\uparrow}$ and $f$ orbital sites, could contain the so-called
}\textit{{poor man's Majorana}}{s.
These MZMs, in particular, cannot be considered topologically protected
as true MZMs, and emerge at the M. Leijnse }\textit{{et
al}}{{} sweet spot\cite{Flensberg_2012(Poor)}, when
the following set of parameters is obeyed: $J=0,$ $\Delta=t$ and
$\varepsilon_{M}=\varepsilon_{\uparrow}=0.$ This scenario can be
observed in the inset panels a)-I of Figs.\ref{fig:Fig3} and \ref{fig:Fig5},
which reveal resonant MZMs, namely, the}\textit{{{}
poor man's Majorana}}{s, with one placed at left and
the other at right of the Kitaev dimer. This statement holds, since
the spectral weights are given by the $\text{DOS}(\uparrow)[\gamma_{A}](0)=1$
and $\text{DOS}(\uparrow)[\gamma_{B}](0)=0$ for the QI {[}inset a)-I
of Fig.\ref{fig:Fig3}(a){]}, while we have $\text{DOS}[\gamma_{1}](0)=0$
and $\text{DOS}[\gamma_{2}](0)\neq0$ for the Kitaev dimer right edge,
as depicted in the inset of Fig.\ref{fig:Fig5}(a). For the aforementioned
case, the nonlocal fermion $\eta=(\gamma_{A}-i\gamma_{2})/\sqrt{2}$
can be made via the linear combination between the resonant MZMs $\gamma_{2}$
and $\gamma_{A}$ localized at right and left of the dimer, respectively\cite{Flensberg_2012(Poor)}.
In this manner, the fermion parity, which is given by the electronic
occupation of $\eta^{\dagger}\eta,$ becomes a feasible quantity for
quantum computing\cite{Flensberg_2012(Poor)}. For the resonant MZMs
$\gamma_{A}$ and $\gamma_{B}$ at left depicted in central panel
of Fig.\ref{fig:Fig3}(a) and the corresponding in Fig.\ref{fig:Fig5}(a),
i.e., the situation off the M. Leijnse }\textit{{et
al}}{{} sweet spot with $\varepsilon_{M}=\Gamma,$
the fragility of these}\textit{{{} poor man's Majoranas}}{{}
becomes evident. In such a case, the spectral weights $\text{DOS}(\uparrow)[\gamma_{A}](0)=1$
and $\text{DOS}(\uparrow)[\gamma_{B}](0)=1$ occur simultaneously
at the QI. Thus, according to Ref.\cite{Flensberg_2012(Poor)}, the
zero-mode dip in the DOS $(-1/\pi)\text{{Im}}\langle\langle\gamma_{1};\gamma_{1}\rangle\rangle_{\omega}$
represents the spill of the MZM $\gamma_{1}$ from the $f$ orbital
site over the QI, being characterized by the $\text{DOS}(\uparrow)[\gamma_{B}](0)=1.$
As aftermath, these two resonant MZMs inevitably introduce an ambiguous
definition for the nonlocal parity qubit, i.e., it could be $\eta=(\gamma_{A}-i\gamma_{2})/\sqrt{2}$
or $\eta=(\gamma_{B}-i\gamma_{2})/\sqrt{2}.$ In this way, the parity
qubit becomes not well-defined and its employment compromised, as
expected, for quantum computing.}

{With this in mind, we back to our sweet spot in Fig.\ref{fig:Fig3},
where the Ising term $J_{\text{{h}}},$ by considering $\varepsilon_{M}=\Gamma,$
siphons off the $\text{DOS}(\uparrow)[\gamma_{A}](0)=1$ {[}Fig.\ref{fig:Fig3}(a){]}
from resonant profile towards one antiresonant with $\text{DOS}(\uparrow)[\gamma_{A}](0)=1$/2
{[}Fig.\ref{fig:Fig3}(c){]}, while it keeps the $\text{DOS}(\uparrow)[\gamma_{B}=\gamma_{\text{{MIQ}}}](0)=1/2$
{[}Fig.\ref{fig:Fig3}(c){]} resonant after such a deviation from
the value $\text{DOS}(\uparrow)[\gamma_{B}](0)=1$ {[}Fig.\ref{fig:Fig3}(a){]}.
As the MZM given by $\gamma_{\text{{MIQ}}}$ exhibits resonant character
and its partner $\gamma_{A}$ is of antiresonant-type, it means that
the QI does not host exactly one MZM. It is an  indication that a
MZM is spilled over the QI from the $f$ orbital site, where the $\text{DOS}[\gamma_{1}](0)=0$
and the resonant $\text{DOS}[\gamma_{2}](0)\neq0$ still persist,
as can be seen in Fig.\ref{fig:Fig5}(b). Thus, such features reinforce
that the MIQ excitation obeys the properties of the }{poor
man's Majorana}\cite{Flensberg_2012(Poor)}. Let
us remember that  in panels a)-I from Figs.\ref{fig:Fig3} and \ref{fig:Fig5},
the left and right resonant MZMs were linearly combined to build $\eta$
when $\varepsilon_{M}=0.$ However, an extrapolation to $\eta=(\gamma_{\text{{MIQ}}}-i\gamma_{2})/\sqrt{2}$
for $\varepsilon_{M}\neq0$ instead, despite being $\gamma_{B}=\gamma_{\text{{MIQ}}}$
resonant and $\gamma_{A}$ not, deserves further investigation, once
Ref.\cite{Flensberg_2012(Poor)} does not cover such a limit. We let
the advances on this particular parity qubit issue to better exploration
elsewhere, since they do not belong to the scope of our current research.
We call attention that our main findings consist of showing one way
of realizing the fractionalization of MZMs, in particular, via the
Ising coupling to a QI site.

{Still concerning the nature of the fractionalized
spectral weight peak given by $\text{DOS}(\uparrow)[\gamma_{\text{{MIQ}}}](0)=1/2$,
we attribute its origin to several modes that added up result into
a half-integer contribution.}{Indeed, each mode,
as we know, due to the QI-leads coupling, becomes a band centered
at the mode in the absence of the leads. As the separation between
the modes (centers of these bands) depends directly on the Ising coupling
$J,$ by decreasing to $J=J_{\text{{h}}}$ (our sweet spot), it favors
the partial merge of the $2S+1$ odd number of bands at zero-energy,
where there is an accumulation point. This is naturally imposed by
the choice of an integer $S$. Consequently, we find the peak $\text{DOS}(\uparrow)[\gamma_{\text{{MIQ}}}](0)=1/2$
and the dip $\text{DOS}(\uparrow)[\gamma_{A}](0)=1/2,$ when we fix}
{$\text{{DOS}}(\uparrow)(0)=1/2$ in Eq.(\ref{eq:NewDOSelectron}). }

{In summary, as demonstrated by us, the MIQ excitation
(the resonant one) and $\gamma_{A}$ (the corresponding partner antiresonant)
at the system left, together with $\gamma_{2}$ at right, are }\textit{{poor
man's Majorana}}{s, since they obey their properties
of partially protection (not topological) introduced by Ref.\cite{Flensberg_2012(Poor)}.
It is worth mentioning that }\textit{{poor man's Majorana}}{s
were recently verified in the experiment of Ref.\cite{Kouwenhoven_2023}
and in our case, the }\textit{{poor man's Majorana}}{s
observed at the QI show fractionalized characteristic, due to the
Ising coupling. At the M. Leijnse }\textit{{et al}}{{}
sweet spot\cite{Flensberg_2012(Poor)}, the Ising term plays no role
and we still have $\text{{DOS}}(\uparrow)(0)=1/2,$ but with the $\text{DOS}(\uparrow)[\gamma_{B}](0)=0$
and a peak in the $\text{DOS}(\uparrow)[\gamma_{A}](0)=1,$ thus reflecting
not fractionalized MZMs. In both the sweet spots, it is capital to
note that we have always $\mathcal{G}_{\text{{Total}}}(0)=\frac{e^{2}}{2h},$
once we adopt the right side of Eq.(\ref{eq:Dup}) as the basis for
the quasiparticle excitations of the electron at the QI {[}Eq.(\ref{eq:DOSelectron}){]}.
This basis is convenient to evaluate the system quantum transport
and elucidates if just one MZM (M. Leijnse }\textit{{et
al}}{{} sweet spot) or half of each from the MZM
couple of quasiparticle excitations at the QI (our sweet spot), is
really contributing to $\mathcal{G}_{\text{{Total}}}(0)=\frac{e^{2}}{2h}.$ }

{Figs.\ref{fig:Fig5}(c) and (d) discuss the ABSs
regime, which can be captured by imposing $|\lambda_{1}|\sim|\lambda_{2}|$
{[}$\Delta\neq0,$ $t\rightarrow0${]} in Eq.(\ref{eq:Htotal}) {[}Eq.(\ref{eq:HMZMs}){]},
as pointed out by Ref.\cite{Nonlocality_Majorana}. It is worth mentioning
that such a scenario corresponds to place the MZMs $\gamma_{1}$ and
$\gamma_{2}$ practically at same TSC edge {[}see Fig.\ref{fig:Fig1}(d){]},
with model parameters marked in Fig.\ref{fig:Fig5}(c), from where
we highlight $\Delta=1.5\Gamma$ and $t=0.01\Gamma.$ Thus, it leads
to a conductance $\mathcal{G}_{\text{{Total}}}(0)=\frac{e^{2}}{2h},$
which is purely from electron tunneling, but counterintuitively, assisted
by Andreev reflection. It means that although $\mathcal{G}_{\text{{LAR}}}(\text{{eV}})\rightarrow0$
through the QI flanked by the leads {[}Eqs.(\ref{eq:LAR}), (\ref{eq:GFforLAR})
and (\ref{eq:GFdd}){]}, the Kitaev dimer, indeed still admits Andreev
reflection, in particular, between the $d_{\uparrow}$ and $f$ orbitals
given by the terms $\Delta d_{\uparrow}f+\text{H.c.}$ {[}Eq.(\ref{eq:HMZMs}),\cite{Flensberg_2012(Poor)}{]},
which then build the quasiparticle electronic state at the QI. As
Eqs.(\ref{eq:NewGtotal}) and (\ref{eq:GMajorana}) also determine
$\mathcal{G}_{\text{{Total}}}(0)=\frac{e^{2}}{2h},$ in Fig.\ref{fig:Fig5}(c),
we evaluate then the DOSs $\text{DOS}(\uparrow)[\gamma_{j}]=-\Gamma\text{{Im}}\langle\langle\gamma_{j};\gamma_{j}\rangle\rangle_{\omega}$
{[}Eq.(\ref{eq:DOSMajorana}){]}, which strongly depend only on the
GFs $\langle\langle d_{\uparrow};d_{\uparrow}^{\dagger}\rangle\rangle_{\omega}$
and $\langle\langle d_{\uparrow}^{\dagger};d_{\uparrow}\rangle\rangle_{\omega}$
{[}Eq.(\ref{eq:GamajGamaj}){]}, since $\langle\langle d_{\uparrow};d_{\uparrow}\rangle\rangle_{\omega}=\langle\langle d_{\uparrow}^{\dagger};d_{\uparrow}^{\dagger}\rangle\rangle_{\omega}\rightarrow0$
{[}Eqs.(\ref{eq:GFforLAR}) and (\ref{eq:GFdd}){]} in the ABS regime.
By looking at Eqs.(\ref{eq.Green dddag}) and (\ref{eq:Green ddagd})
in the Appendix A, we verify that $\langle\langle d_{\uparrow};d_{\uparrow}^{\dagger}\rangle\rangle_{\omega}$
and $\langle\langle d_{\uparrow}^{\dagger};d_{\uparrow}\rangle\rangle_{\omega}$
have a dependence on $\Delta\langle\langle f^{\dagger}|m\rangle\langle m|;d_{\uparrow}^{\dagger}\rangle\rangle_{\omega}$
and $\Delta\langle\langle f|m\rangle\langle m|;d_{\uparrow}\rangle\rangle_{\omega},$
which are modulated by the SC pairing $\Delta$ and GFs associated
to $\Delta d_{\uparrow}f+\text{H.c.}$ for the Andreev process. Such
features establish the Andreev reflection between the QI and $f,$
while $\mathcal{G}_{\text{{LAR}}}(\text{{eV}})\rightarrow0$ is evaluated
at the interface QI-leads. Additionally, we should highlight that
similar analysis in observing the ABSs regime, in particular, within
the effective model of Eq.(\ref{eq:Htotal}) but without the Ising
term, was already performed by some of us via the analogous Eq.(20)
of Ref.\cite{Ricco_2019} and now, we extend it to the current Hamiltonian.
As consequences, Fig.\ref{fig:Fig5}(c) also exhibits the DOSs siphon
off to $1/2$ for $\gamma_{A}$ and $\gamma_{B},$ which distinct
from Figs.\ref{fig:Fig5}(a) and (b) with $\Delta=t,$ are identically
resonant. In Fig.\ref{fig:Fig5}(d), a pair of ABSs manifests too.
To conclude, $\mathcal{G}_{\text{{Total}}}(0)=\frac{e^{2}}{2h}$ could
be reproduced by both the }\textit{{poor man's Majorana}}{s
and ABSs regimes. }

\section{Conclusions}

We found that the fractionalization of regular MZMs becomes a feasible
task once an integer large spin $S$ is exchange coupled to a quantum
impurity, in particular when it acts as the new edge of a finite TSC
in 1D. A counterintuitive regime arises due to a sweet value for the
Ising coupling, {which is capable of localizing a
fractionalized MZM.} We introduce such an excitation as the called
Majorana-Ising-type quasiparticle (MIQ). As aftermath, we report the
emergence of one MZM with the maximum spectral weight reduced by half
and exhibiting resonant character. In contrast, the other MZM mode
in the QI does not localize around zero-energy, but shows the same
spectral weight of the resonant MZM via an antiresonant profile. Interestingly
enough, due to the localization of the MIQ, half of the quantum conductance
is made essentially by the normal electronic contribution, while that
from {the Andreev reflection is totally lacking between
the QI and leads.} This behavior differs from that observed in perfectly
infinite TSC wires, in which one MZM localizes at QI site with maximum
spectral weight given by unit and with electronic and Andreev conductances
equally split at zero-bias. Therefore, our proposal points out a manner
to induce, within a more realistic perspective from an experimental
point of view, a quantum state at the edge of a short TSC in 1D. {Additionally,
our MIQ is demonstrated to be }\textit{{a poor man's
Majorana}}{\cite{Flensberg_2012(Poor),Kouwenhoven_2023}.}

\section{Acknowledgments}

We thank the Brazilian funding agencies CNPq (Grants. Nr. 302887/2020-2,
308410/2018-1, 311980/2021-0, 305668/2018-8 and 308695/2021-6), Coordenação
de Aperfeiçoamento de Pessoal de Nível Superior - Brasil (CAPES) --
Finance Code 001 and FAPERJ process Nr. 210 355/2018. LSR and IAS
acknowledge the support from Icelandic Research Fund (Rannis), projects
No. 163082-051 and ``Hybrid polaritonics''. LSR thanks ACS and Unesp
for their hospitality.\linebreak{}

\appendix

\section{{Green's functions for the system left side}}

As the GFs in the presence of the large spin obey the notation $\langle\langle A_{\sigma};B_{\sigma}\rangle\rangle_{\omega}=\sum_{m}\langle\langle A_{\sigma}|m\rangle\langle m|;B_{\sigma}\rangle\rangle_{\omega}${\citep{Zubarev-1960}},
here we make explicit the details in the EOM approach to find the
elements of type $\langle\langle A_{\sigma}|m\rangle\langle m|;B_{\sigma}\rangle\rangle_{\omega}.$
In what follows, we have
\begin{eqnarray}
 &  & (\omega^{+}-\varepsilon_{\uparrow}-\frac{Jm}{2}+i\Gamma)\langle\langle d_{\uparrow}|m\rangle\langle m|;d_{\uparrow}^{\dagger}\rangle\rangle_{\omega}\nonumber \\
 & = & \frac{1}{2S+1}-t\langle\langle f|m\rangle\langle m|;d_{\uparrow}^{\dagger}\rangle\rangle_{\omega}-\Delta\langle\langle f^{\dagger}|m\rangle\langle m|;d_{\uparrow}^{\dagger}\rangle\rangle_{\omega},\nonumber \\
\label{eq.Green dddag}
\end{eqnarray}
\begin{eqnarray}
 &  & (\omega^{+}+\varepsilon_{\uparrow}+\frac{Jm}{2}+i\Gamma)\langle\langle d_{\uparrow}^{\dagger}|m\rangle\langle m|;d_{\uparrow}\rangle\rangle_{\omega}\nonumber \\
 & = & \frac{1}{2S+1}+t\langle\langle f^{\dagger}|m\rangle\langle m|;d_{\uparrow}\rangle\rangle_{\omega}+\Delta\langle\langle f|m\rangle\langle m|;d_{\uparrow}\rangle\rangle_{\omega},\nonumber \\
\label{eq:Green ddagd}
\end{eqnarray}
the other two terms are given by
\begin{eqnarray}
 &  & (\omega^{+}+\varepsilon_{\uparrow}+\frac{Jm}{2}+i\Gamma)\langle\langle d_{\uparrow}^{\dagger}|m\rangle\langle m|;d_{\uparrow}^{\dagger}\rangle\rangle_{\omega}\nonumber \\
 & = & t\langle\langle f^{\dagger}|m\rangle\langle m|;d_{\uparrow}^{\dagger}\rangle\rangle_{\omega}+\Delta\langle\langle f|m\rangle\langle m|;d_{\uparrow}^{\dagger}\rangle\rangle_{\omega}
\end{eqnarray}
and
\begin{eqnarray}
 &  & (\omega^{+}-\varepsilon_{\uparrow}-\frac{Jm}{2}+i\Gamma)\langle\langle d_{\uparrow}|m\rangle\langle m|;d_{\uparrow}\rangle\rangle_{\omega}\nonumber \\
 & = & -t\langle\langle f|m\rangle\langle m|;d_{\uparrow}\rangle\rangle_{\omega}-\Delta\langle\langle f^{\dagger}|m\rangle\langle m|;d_{\uparrow}\rangle\rangle_{\omega}
\end{eqnarray}
And finally the last two GFs associated with the $f$ site is
\begin{eqnarray}
\langle\langle f|m\rangle\langle m|;d_{\uparrow}\rangle\rangle_{\omega} & = & \frac{-t}{(\omega^{+}-\varepsilon_{M})}\langle\langle d_{\uparrow}|m\rangle\langle m|;d_{\uparrow}\rangle\rangle_{\omega}\nonumber \\
 & + & \frac{\Delta}{(\omega^{+}-\varepsilon_{M})}\langle\langle d_{\uparrow}^{\dagger}|m\rangle\langle m|;d_{\uparrow}\rangle\rangle_{\omega}\nonumber \\
\end{eqnarray}
and
\begin{eqnarray}
\langle\langle f^{\dagger}|m\rangle\langle m|;d_{\uparrow}\rangle\rangle_{\omega} & = & \frac{t}{(\omega^{+}+\varepsilon_{M})}\langle\langle d_{\uparrow}^{\dagger}|m\rangle\langle m|;d_{\uparrow}\rangle\rangle_{\omega}\nonumber \\
 & - & \frac{\Delta}{(\omega^{+}+\varepsilon_{M})}\langle\langle d_{\uparrow}|m\rangle\langle m|;d_{\uparrow}\rangle\rangle_{\omega}.\nonumber \\
\end{eqnarray}
With this group of GFs we can determine the complete description of
the QI.

\section{{Quantum transport formalism}}

{Based on the quantum transport Keldysh formalism
of Ref.\cite{Zarbo-2022}, we wrap up here a summary of steps in deriving
Eqs.(\ref{eq:ET}), (\ref{eq:LAR}) and (\ref{eq:DOSelectron}), which
assumes the subgap regime $\left|\text{{eV}}\right|\ll\Delta_{\text{{SC}}}\rightarrow\infty$
(infinite superconducting gap standard approximation) and wide-band
limit characterized by an electron-hole symmetry in the QI-leads coupling,
which is given by $\Gamma$ {[}see the main text below Eq.(\ref{eq:Htotal}){]}.
As a result, we have as the total current $I_{\alpha}$ at the metallic
lead $\alpha$ the following
\begin{eqnarray}
I_{\alpha} & = & I_{\alpha}^{\text{ET}}+I_{\alpha}^{\text{LAR}}+I_{\alpha}^{\text{CAR}},\label{eq:Current1}
\end{eqnarray}
where }

{
\begin{eqnarray}
I_{\alpha}^{\text{ET}} & = & \frac{e}{h}\int d\varepsilon\tau_{\alpha\bar{\alpha}}^{\text{ET}}(\varepsilon)\left[f_{\alpha}^{e}(\varepsilon)-f_{\bar{\alpha}}^{e}(\varepsilon)\right],\label{eq:CurrentET}
\end{eqnarray}
\begin{eqnarray}
I_{\alpha}^{\text{CAR}} & = & \frac{e}{h}\int d\varepsilon\tau_{\alpha\bar{\alpha}}^{\text{CAR}}(\varepsilon)\left[f_{\alpha}^{e}(\varepsilon)-f_{\bar{\alpha}}^{h}(\varepsilon)\right]\label{eq:CurrentCAR}
\end{eqnarray}
and
\begin{eqnarray}
I_{\alpha}^{\text{LAR}} & = & \frac{e}{h}\int d\varepsilon\tau_{\alpha\alpha}^{\text{LAR}}(\varepsilon)\left[f_{\alpha}^{e}(\varepsilon)-f_{\alpha}^{h}(\varepsilon)\right],\label{eq:CurrentLAR}
\end{eqnarray}
where $I_{\alpha}^{\text{ET}}$ and $I_{\alpha}^{\text{CAR}}$ refer
to the currents for the electron tunneling (ET) and crossed Andreev
reflection (CAR) between the QI and the lead $\alpha,$ but with occupation
probabilities of an electron $f_{\bar{\alpha}}^{e}(\varepsilon)$
and hole $f_{\bar{\alpha}}^{h}(\varepsilon)$ states at lead $\bar{\alpha},$
respectively, where $f_{\alpha}^{j}(\varepsilon)$ stands for the
Fermi distribution at lead $\alpha$ and $j=e(h)$ for the electron
(hole) quasiparticle. For the local Andreev reflection (LAR) $I_{\alpha}^{\text{LAR}},$
the hole emission is into the same terminal $\alpha,$ once it depends
on $f_{\alpha}^{h}(\varepsilon).$}

{As the total current should conserve, the Kirchhoff's
law holds $I_{\alpha}+I_{\bar{\alpha}}+I_{S}=0.$ Additionally, the
following assumptions are performed: $\mu_{\text{{Source}}}=-\mu_{\text{{Drain}}}=\text{{eV}}/2$
and the TSC is supposed to be grounded (null chemical potential $\mu_{\text{SC}}=0).$
The former implies in $f_{\alpha}^{e}(\varepsilon)=f_{\bar{\alpha}}^{h}(\varepsilon)$
and consequently $I_{\alpha}^{\text{CAR}}=0$ from Eq.(\ref{eq:CurrentCAR}),
while the latter gives $I_{S}=0$ and finally $I_{\alpha}=-I_{\bar{\alpha}}.$
It means that the current only changes the sign from terminal $\alpha$
to $\bar{\alpha},$ being the conductance $\mathcal{G}_{\text{{Total}}}$
lead independent: }

{{}
\begin{equation}
\mathcal{G}_{\text{{Total}}}=\frac{dI_{\alpha}}{d\text{{V}}}=\frac{dI_{\alpha}^{\text{ET}}}{d\text{{V}}}+\frac{dI_{\alpha}^{\text{LAR}}}{d\text{{V}}}=\mathcal{G}_{\text{{ET}}}+\mathcal{G}_{\text{{LAR}}},\label{eq:Gtotal2}
\end{equation}
with }

{
\begin{eqnarray}
\frac{dI_{\alpha}^{\text{ET}}}{d\text{{V}}} & = & \frac{e^{2}}{2h}\frac{1}{T}\int d\varepsilon\tau_{\alpha\bar{\alpha}'}^{\text{ET}}(\varepsilon)\{f_{\alpha}^{e}(\varepsilon)[1-f_{\alpha}^{e}(\varepsilon)]\nonumber \\
 & + & f_{\bar{\alpha}}^{e}(\varepsilon)[1-f_{\bar{\alpha}}^{e}(\varepsilon)]\},
\end{eqnarray}
and
\begin{eqnarray}
\frac{dI_{\alpha}^{\text{LAR}}}{d\text{{V}}} & = & \frac{e^{2}}{2h}\frac{1}{T}\int d\varepsilon\tau_{\alpha\alpha}^{\text{LAR}}(\varepsilon)\{f_{\alpha}^{e}(\varepsilon)[1-f_{\alpha}^{e}(\varepsilon)]\nonumber \\
 & + & f_{\bar{\alpha}}^{e}(\varepsilon)[1-f_{\bar{\alpha}}^{e}(\varepsilon)]\},
\end{eqnarray}
where we employed the identity $f_{\alpha}^{e}(\varepsilon)=f_{\bar{\alpha}}^{h}(\varepsilon)$
again and }

{{}
\begin{equation}
\frac{\partial f_{\alpha(\bar{\alpha})}^{e}(\varepsilon)}{\partial\text{{V}}}=\pm\frac{e}{2T}f_{\alpha(\bar{\alpha})}^{e}(\varepsilon)[1-f_{\alpha(\bar{\alpha})}^{e}(\varepsilon)],\label{eq:Derivative}
\end{equation}
with $k_{B}=1,$ $f_{\alpha}^{e}(\varepsilon)=f(\varepsilon-\text{{eV}}/2)$,
$f_{\bar{\alpha}}^{e}(\varepsilon)=f(\varepsilon+\text{{eV}}/2)$
and $f(x)=1/(1+e^{x/T})$. }

{As $\frac{1}{T}f_{\alpha(\bar{\alpha})}^{e}(\varepsilon)[1-f_{\alpha(\bar{\alpha})}^{e}(\varepsilon)])=\left(-\frac{\partial f_{\alpha(\bar{\alpha})}^{e}(\varepsilon)}{\partial\varepsilon}\right)\rightarrow\delta(\varepsilon\mp\text{{eV}}/2)$
when $T\rightarrow0\text{{K},}$ then we deduce Eqs.(\ref{eq:ET})
and (\ref{eq:LAR}). To conclude, we should remember that the Keldysh
formalism of Ref.\cite{Zarbo-2022} also ensures }

{
\begin{equation}
I_{\alpha}=\frac{e}{h}\int d\varepsilon(-\Gamma\text{{Im}}\langle\langle d_{\uparrow};d_{\uparrow}^{\dagger}\rangle\rangle_{\omega})\left[f_{\alpha}^{e}(\varepsilon)-f_{\bar{\alpha}}^{e}(\varepsilon)\right].\label{eq:Current2}
\end{equation}
}

{Therefore, by comparing Eqs.(\ref{eq:Current2})
and (\ref{eq:Current1}), together with Eqs.(\ref{eq:CurrentET})
and (\ref{eq:CurrentLAR}), we finally determine Eq.(\ref{eq:DOSelectron}).}

\section{{Green's functions for the system right side}}

{Here we show the GFs for the $f$ orbital site and
the MZMs $\gamma_{1}$ and $\gamma_{2}$ of the Kitaev dimer right
side {[}Eq.(\ref{eq:HMZMs}){]}. These quantities are important for
Fig.\ref{fig:Fig5} and make explicit the }\textit{{poor
man's Majorana}}{s\cite{Flensberg_2012(Poor),Kouwenhoven_2023}
and ABSs\cite{Nonlocality_Majorana} regimes in our system. As we
know that $\gamma_{1}=\frac{1}{\sqrt{2}}(f^{\dagger}+f)$ and $\gamma_{2}=\frac{i}{\sqrt{2}}(f^{\dagger}-f),$
we naturally find the GF}

{{}
\begin{eqnarray}
\langle\langle\gamma_{j};\gamma_{j}\rangle\rangle_{\omega} & = & \frac{1}{2}\left[\langle\langle f;f^{\dagger}\rangle\rangle_{\omega}+\langle\langle f^{\dagger};f\rangle\rangle_{\omega}\right.\nonumber \\
 & + & \left.\epsilon\left(\langle\langle f^{\dagger};f^{\dagger}\rangle\rangle_{\omega}+\langle\langle f;f\rangle\rangle_{\omega}\right)\right],
\end{eqnarray}
where $j=(1,2)$ corresponds to $\epsilon=(+1,-1)$, respectively.
By applying Eq.(\ref{eq:EOM}) for the EOM approach, we obtain }

{
\begin{eqnarray}
\langle\langle f;f^{\dagger}\rangle\rangle_{\omega} & = & \frac{1}{\left(2S+1\right)}\sum_{m}\frac{1}{\left(\omega^{+}-\varepsilon_{M}-\Sigma_{+M}\right)},
\end{eqnarray}
\begin{eqnarray}
\langle\langle f^{\dagger};f\rangle\rangle_{\omega} & = & \frac{1}{\left(2S+1\right)}\sum_{m}\frac{1}{\left(\omega^{+}+\varepsilon_{M}-\Sigma_{-M}\right)},
\end{eqnarray}
\begin{eqnarray}
\langle\langle f^{\dagger};f^{\dagger}\rangle\rangle_{\omega} & = & \frac{1}{\left(2S+1\right)}\sum_{m}\frac{2\Delta tK_{-M}}{\left(\omega^{+}+\varepsilon_{M}-\Sigma_{-M}\right)}
\end{eqnarray}
and
\begin{eqnarray}
\langle\langle f;f\rangle\rangle_{\omega} & = & \frac{1}{\left(2S+1\right)}\sum_{m}\frac{2\Delta tK_{+M}}{\left(\omega^{+}-\varepsilon_{M}-\Sigma_{+M}\right)}.
\end{eqnarray}
with $\Sigma_{\pm M}=K_{C}^{\pm}+\left(2\Delta t\right)^{2}K_{D}K_{\pm M}$
being the self-energy due to the interaction of the $f$ site with
the QI, which can be expressed in terms of the defined quantities
\begin{eqnarray}
K_{+M} & = & \frac{K_{D}}{\left(\omega^{+}+\varepsilon_{M}-K_{C}^{-}\right)},
\end{eqnarray}
}\\
{
\begin{eqnarray}
K_{-M} & = & \frac{K_{D}}{\left(\omega^{+}-\varepsilon_{M}-K_{C}^{+}\right)},
\end{eqnarray}
}\\
{
\begin{eqnarray}
K_{D} & = & \frac{\left(\omega+i\eta^{+}\right)+i\Gamma}{\left(\omega+i\Gamma+i\eta^{+}\right)^{2}-\varepsilon_{\uparrow}^{2}-Jm\left[\varepsilon_{\uparrow}+\frac{Jm}{4}\right]},
\end{eqnarray}
and }

{{}
\begin{eqnarray}
K_{C}^{\pm} & = & \left[\frac{\left(\omega+i\eta^{+}\right)\left(t^{2}+\Delta^{2}\right)\pm\varepsilon_{\uparrow}\left(t^{2}-\Delta^{2}\right)}{\left(\omega+i\Gamma+i\eta^{+}\right)^{2}-\varepsilon_{\uparrow}^{2}-Jm\left[\varepsilon_{\uparrow}+\frac{Jm}{4}\right]}\right]\nonumber \\
 & + & \left[\frac{\pm\frac{Jm}{2}\left(t^{2}-\Delta^{2}\right)+i\Gamma\left(t^{2}+\Delta^{2}\right)}{\left(\omega+i\Gamma+i\eta^{+}\right)^{2}-\varepsilon_{\uparrow}-Jm\left[\varepsilon_{\uparrow}+\frac{Jm}{4}\right]}\right].
\end{eqnarray}
}

\bibliographystyle{apsrev4-2}

\end{document}